\documentclass{JHEP3}
\usepackage{amsmath}
\usepackage{graphicx}
\usepackage{cite}

\allowdisplaybreaks[1]

\makeatletter

\@addtoreset{equation}{section}

\newcommand\Ref[1]     {Ref.\,\cite{#1}}

\newcommand\refr[1]      {ref.\,\cite{#1}}
\newcommand\refrs[1]    {refs.\,\cite{#1}}

\newcommand\eqn[1]     {eq.\,(\ref{#1})}
\newcommand\eqns[2]    {eqs.\,(\ref{#1}) and~(\ref{#2})}
\newcommand\eqnss[2]   {eqs.\,(\ref{#1})--(\ref{#2})}

\newcommand\fig[1]     {fig.\,{\ref{#1}}}

\newcommand\Sect[1]    {Sect.\,{\ref{#1}}}

\newcommand\sect[1]    {sect.\,{\ref{#1}}}

\newcommand\appx[1]     {appendix~\ref{#1}}

\newcommand\nt         {\notag}

\def\beq{\begin{equation}}
\def\eeq{\end{equation}}
\def\bsp#1\esp{\begin{split}#1\end{split}}
\def\bal#1\eal{\begin{align}#1\end{align}}

\newcommand\bom[1]     {{\mbox{\boldmath $#1$}}}

\newcommand\as         {\ensuremath{\alpha_{\mathrm{s}}}}

\newcommand\pref    {8\pi\as\mu^{2\eps}}

\newcommand{\CF}       {C_{\mathrm{F}}}
\newcommand{\CA}       {C_{\mathrm{A}}}
\newcommand{\TR}       {T_{\mathrm{R}}}
\newcommand{\Nc}       {N_{\mathrm{c}}}
\newcommand{\Nf}       {n_{\mathrm{f}}}

\newcommand{\bT}       {\bom{T}}

\newcommand\qb         {{\bar q}}
\newcommand\msbar      {\ensuremath{{\overline {\rm MS}}}}

\newcommand\muF[1]     {\ensuremath{\mu_F^{#1}}}
\newcommand\e          {{\mathrm e}}

\newcommand\Oe[1]      {\ensuremath{\mathrm O(\eps^{#1})}}
\newcommand{\eps}      {\varepsilon}

\newcommand\Li         {\mathop{\mathrm{Li}}\nolimits}

\newcommand\shalf	{{\textstyle \frac{1}{2}}}

\newcommand{\finite}   {{\cal F}\!in}        

\newcommand{\PS}[1]    {\rd\phi_{#1}}
\newcommand{\rd}{{\mathrm{d}}}
\newcommand\tsig[1]    {\sigma^{\mathrm{#1}}}
\newcommand\dsig[1]    {\rd\sigma^{{\rm #1}}}
\newcommand\dsiga[2]   {\rd\sigma^{{\rm #1,A}_{\scriptscriptstyle #2}}}

\newcommand\la         {\langle}
\newcommand\ra         {\rangle}

\newcommand{\cA}       {{\cal A}}
\newcommand{\cM}       {{\cal M}}
\newcommand\SME[3]     {|{\cal M}_{#1}^{(#2)}{(#3)}|^2}

\newcommand\bra[3]     {\la {\cal M}_{#1}^{#2}#3|}
\newcommand\ket[3]     {|{\cal M}_{#1}^{#2}#3\ra}

\newcommand{\mom}[1]   {\{p\}^{#1}}
\newcommand{\momt}[1]   {\{\ti{p}\}^{#1}}

\newcommand{\momab}[1]    {\{p\}_{#1};p_a,p_b}

\newcommand{\momabt}[2]    {\{\ti{p}\}_{#1}^{#2};p_a,p_b}
\newcommand{\momabtx}[4]    {\{\ti{p}\}_{#1}^{#2};#3 p_a,#4 p_b}
\newcommand{\momabtt}[4]     {\{\ti{p}\}_{#1}^{#2};\ti{p}_{#3},p_{#4}}

\newcommand{\bA}[1]    {\bom{\mathrm A}_{#1}}

\newcommand{\bC}[1]    {\bom{\mathrm C}_{#1}}
\newcommand{\bS}[1]    {\bom{\mathrm S}_{#1}}

\newcommand{\bSCS}[1]  {\bom{\mathrm C}\kern-2pt\bom{\mathrm S}_{#1}}

\def\hP{\hat{P}}

\newcommand{\calS}     {{\cal S}}

\newcommand{\cC}[2]    {{\cal C}_{#1}^{#2}}
\newcommand{\cS}[2]    {{\cal S}_{#1}^{#2}}

\newcommand{\cSCS}[1]  {{\cal C}\kern-2pt{\cal S}_{#1}^{~}}

\newcommand{\IcC}[2]   {{\mathrm C}_{#1}^{#2}}
\newcommand{\IcS}[2]   {{\mathrm S}_{#1}^{#2}}
\newcommand{\TcS}[2]   {\ti{\mathrm S}_{#1}^{#2}}
\newcommand{\IcCS}[1]  {\mathrm{C\!S}^{#1}}

\newcommand{\bI}       {\bom{I}}

\newcommand{\ti}[1]    {\tilde{#1}}
\newcommand{\wti}[1]   {\widetilde{\,#1\,}}

\newcommand\zt[1]      {\tilde{z}_{#1}}
\newcommand\xt[1]      {\tilde{x}_{#1}}

\newcommand\kT      {k_{\perp}}
\newcommand\kTt[1]     {\tilde{k}_{\perp,#1}}
\newcommand\kTtm[2]    {\tilde{k}_{\perp,#1}^{#2}}
\newcommand\Y[2]   {Y_{\ti{#1}\ti{#2},Q}}

\newcommand\FS         {{\scriptscriptstyle\rm F\!.S\!.}}

\newcommand{\ts}[1]{\textstyle{#1}}

\title{Subtraction with hadronic initial states at NLO:\\ an NNLO-compatible scheme}

\author{G\'abor Somogyi \\
Institute for Theoretical Physics, University of Z\"urich\\ 
Winterthurerstrasse 190, CH-8057 Z\"urich, Switzerland\\
E-mail: \email{sgabi@physik.unizh.ch}}

\abstract{We present an NNLO-compatible subtraction scheme for computing QCD jet cross sections of hadron-initiated processes at NLO accuracy. The scheme is constructed specifically with those complications in mind, that emerge when extending the subtraction algorithm to next-to-next-to-leading order. It is therefore possible to embed the present scheme in a full NNLO computation without any modifications.}

\preprint{arXiv:0903.1218 [hep-ph]\\ ZU-TH 03/09}

\begin{document}

\renewcommand{\thefootnote}{\fnsymbol{footnote}}



\section{Introduction}
\label{sec:intro}

Exploiting the full physics potential of the LHC puts strong demands on the precise theoretical understanding of QCD. In particular, accurate predictions of both signal and background cross sections require the computation of radiative corrections at least at the next-to-leading order (NLO) accuracy. In some cases however, NLO accuracy is not yet satisfactory and one would like to be able to compute perturbative corrections beyond NLO. The physical situations when this happens have been discussed extensively in the literature \cite{Glover:2002gz}. However, the straightforward (albeit toilsome) application of QCD perturbation theory to compute higher order corrections runs into the following problem. The finite higher order corrections are sums of several pieces which are separately infrared (IR) divergent in $d=4$ spacetime dimensions. Therefore the evaluation of phase space integrals beyond LO is not straightforward because it involves IR singularities that have to be consistently treated before any numerical computation may be performed. At NLO accuracy the IR divergences present in the intermediate stages of calculation may be handled using a {\em subtraction algorithm} exploiting the fact that the structure of kinematical singularities of QCD matrix elements is universal. Thus it is possible to construct process-independent counterterms which regularize the one-loop (or virtual) corrections and real-emission phase space integrals simultaneously \cite{Frixione:1995ms,Nagy:1996bz,Frixione:1997np,Catani:1996vz}.

In recent years a lot of effort has been devoted to the extension of the subtraction method to the computation of radiative corrections at the NNLO accuracy \cite{Weinzierl:2003fx,Weinzierl:2003ra,GehrmannDeRidder:2004tv,GehrmannDeRidder:2004xe,Frixione:2004is,GehrmannDeRidder:2005hi,GehrmannDeRidder:2005aw,Somogyi:2005xz,GehrmannDeRidder:2005cm,Weinzierl:2006ij,Somogyi:2006da,Somogyi:2006db,GehrmannDeRidder:2007jk,Weinzierl:2008iv}. However, extending existing NLO subtraction schemes to NNLO accuracy faces the following problem. Consider the NNLO correction to a generic production cross section (for the sake of simplicity, below we will write formulae appropriate to $e^+e^-\to m$ jet production). It is the sum of three pieces, the doubly-real emission, the real-virtual and the doubly-virtual terms
\beq
\tsig{NNLO} = \int_{m+2} \dsig{RR}_{m+2} J_{m+2} 
	+ \int_{m+1} \dsig{RV}_{m+1} J_{m+1} + \int_m \dsig{VV}_m J_m\,.
\label{eq:sigmaNNLO}
\eeq
The above expression is formal, because $\dsig{RR}_{m+2}$ is IR singular in the one- and two-parton unresolved regions of phase space, $\dsig{RV}_{m+1}$ has explicit $\eps$-poles {\em and} is singular in the one-parton unresolved regions of phase space, while $\dsig{VV}_m$ again has poles in $\eps$. Let us concentrate on the singly-unresolved singularities of $\dsig{RR}_{m+2}$ and the $\eps$-poles of $\dsig{RV}_{m+1}$: these are just the divergences which are present in an NLO computation of $m+1$ jet production. Thus it is a natural first step to try and apply a standard NLO subtraction between the first two terms in \eqn{eq:sigmaNNLO} above to obtain
\beq
\bsp
\tsig{NNLO} &= \int_{m+2} \bigg[\dsig{RR}_{m+2} J_{m+2} 
	- \dsiga{RR}{1}_{m+2} J_{m+1}\bigg]
	+ \int_{m+1} \bigg[\dsig{RV}_{m+1} + \int_1 \dsiga{RR}{1}_{m+2}\bigg] J_{m+1} 
\\
	&+ \int_m \dsig{VV}_m J_m\,,
\esp
\label{eq:sigmaNNLOsub1}
\eeq
where $\dsiga{RR}{1}_{m+2}$ is just the NLO approximate cross section for $m+1$ jet production and $\int_1 \dsiga{RR}{1}_{m+2}$ is its integral over the phase space of one unresolved parton. 

The $m+1$-parton integral in \eqn{eq:sigmaNNLOsub1} above is now finite in $\eps$ by construction, but this is of course not the end of the story yet, as it is still singular in the one-parton unresolved regions of phase space. These singularities would be screened by the jet function in an NLO computation, but this is no longer the case at NNLO. Thus, we need to further subtract suitable approximate cross sections which regularize the singly-unresolved limits of $\dsig{RV}_{m+1}$ and $\int_1 \dsiga{RR}{1}_{m+2}$ respectively. The construction of an approximate cross section to the real-virtual piece is simple enough starting from the universal collinear and soft factorization formulae for 1-loop squared matrix elements \cite{Somogyi:2006db}. But building an approximate cross section for $\int_1 \dsiga{RR}{1}_{m+2}$ may run into the problem that this piece need not obey universal factorization in the collinear limit. Indeed $\int_1 \dsiga{RR}{1}_{m+2}$ is given by an expression of the form
\beq
\int_1 \dsiga{RR}{1}_{m+2} = \dsig{R}_{m+1}\otimes \bI(\eps)\,,
\label{eq:I1dsigRR}
\eeq
where $\dsig{R}_{m+1}$ is the Born cross section for the production of $m+1$ final-state partons and $\bI(\eps)$ is an operator in colour space with universal pole part
\beq
\bI(\eps) = -\frac{\as}{2\pi}\frac{1}{\Gamma(1-\eps)}
	\sum_i\left[\sum_{k\ne i}\frac{1}{\eps^2}
	\left(\frac{4\pi\mu^2}{s_{ik}}\right)^\eps\bT_i\bT_k
	-\frac{\gamma_i}{\eps}\right] + \Oe{0}\,,
\label{eq:Ieps}
\eeq
the sums running over all (initial- and final-state) external QCD partons.
However due to coherent soft gluon radiation from unresolved partons, only the sum 
\beq
\bra{m+1}{(0)}{}\bT_k\bT_i\ket{m+1}{(0)}{} 
	+ \bra{m+1}{(0)}{}\bT_k\bT_r\ket{m+1}{(0)}{}
\eeq
factorizes in the $p_i || p_r$ collinear limit \cite{Somogyi:2006db} ($\bT_{ir}=\bT_i+\bT_r$)
\beq
\bC{ir}\Big(\bra{m+1}{(0)}{}\bT_k\bT_i\ket{m+1}{(0)}{} 
	+ \bra{m+1}{(0)}{}\bT_k\bT_r\ket{m+1}{(0)}{}\Big) \propto
	\frac{1}{s_{ir}}\bra{m}{(0)}{}\bT_{k}\bT_{ir}\hP_{ir}\ket{m}{(0)}{}\,.
\label{eq:CirTkTiTkTr}
\eeq
This factorization is violated by the factors of $s_{ik}^{-\eps}/\eps^2$ that appear in \eqns{eq:I1dsigRR}{eq:Ieps} at $\Oe{0}$. This was noticed in \refr{GehrmannDeRidder:2007jk} too, where it was initially claimed that the terms that violate factorization give a vanishing contribution after integration. However it was later realized that this is not the case, as pointed out by \refr{Weinzierl:2008iv} first. Thus we are lead to define a new NLO subtraction scheme that specifically deals with this issue. We note in passing that it is also necessary that the {\em unintegrated} singly-unresolved approximate cross section have universal collinear limits, and this is not guaranteed by QCD factorization properties. (The approximate cross section we define in this paper will be easily seen to obey universal collinear factorization however.) 

For processes with only colourless partons in the initial state, such a scheme was presented in \refr{Somogyi:2006cz}, and in fact this subtraction scheme can be extended to NNLO accuracy \cite{Somogyi:2006da,Somogyi:2006db} without any modifications.

In the present paper, we construct an NLO subtraction scheme for processes with hadronic initial states, that may be extended to NNLO accuracy. The steps needed to set up this subtraction scheme are the same as in \refr{Somogyi:2006cz}. Starting from the known (tree-level) IR factorization formulae for singly-unresolved emission, we first write all relevant formulae in such a way that their overlap structure is disentangled, and multiple subtraction avoided. Then we carefully define extensions of the formulae, so that they are unambiguously defined away from the IR limits. As it turns out, the `minimal' extension of the limit formulae as in \refr{Somogyi:2006cz} {\em does not} lead to a satisfactory approximate cross section in the present case, and we are forced to consider a `non-minimal' extension. Nevertheless, the structure of our final results will be seen to parallel those of \refr{Somogyi:2006cz} closely, thus the present scheme can be viewed as an extension of the algorithm of that paper to processes with hadrons in the initial state.

The outline of the paper is as follows. In \sect{sec:subtraction} we recall the structure of hadronic NLO cross sections and the implementation of the subtraction procedure in some detail. In \sect{sec:notation} we set up some notation. \Sect{sec:fact} recalls the universal factorization properties of QCD squared matrix elements in the collinear and soft limits. We present the new subtraction terms in \sect{sec:subterms}. In \sect{sec:subimp} we discuss the implementation of our subtraction algorithm and perform the integration of the approximate cross section over the phase space of the unresolved parton. Finally, in \sect{sec:conclusions} we give our conclusions.


\section{The subtraction method at NLO}
\label{sec:subtraction}

%
%

\subsection{QCD jet cross sections at NLO}
\label{ssec:QCDxsecNLO}

Let us consider the cross section of some hadron-initiated process at NLO accuracy, 
\beq
\bsp
\sigma(p_A,p_B) &= 
	\sum_{a,b}\int_0^1\rd\eta_a\, f_{a/A}(\eta_a,\muF{2})
	\int_0^1\rd\eta_b\, f_{b/B}(\eta_b,\muF{2})	
	\left[\tsig{LO}_{ab}(p_a,p_b) 
	+ \tsig{NLO}_{ab}(p_a,p_b;\muF{2})\right]\,,
\label{eq:xsec}
\esp
\eeq
where $p_a \equiv \eta_a\, p_A$, $p_b\equiv \eta_b\, p_B$, the subscripts on the cross sections denote the flavour of the incoming partons while $f_{a/A}$ and $f_{b/B}$ are the parton densities of the two incoming hadrons, $A$ and $B$. Assuming an $m$-jet quantity, the leading order parton-level cross section is given by the integral of the fully differential Born cross section $\dsig{B}_{ab}$ over the available $m$-parton phase space defined by the jet function $J_m$
\beq
\tsig{LO}_{ab}(p_a,p_b) = 
	\int_m \dsig{B}_{ab}(p_a,p_b)J_m\,.
\label{eq:tsigLO}
\eeq
The NLO contribution is a sum of three pieces, the real-emission and virtual corrections and the collinear subtraction term
\beq
\tsig{NLO}_{ab}(p_a,p_b;\muF{2}) = 
	\int_{m+1} \dsig{R}_{ab}(p_a,p_b) J_{m+1}
	+ \int_m \dsig{V}_{ab}(p_a,p_b) J_m
	+ \int_m \dsig{C}_{ab}(p_a,p_b;\muF{2}) J_m\,.
\label{eq:tsigNLO}
\eeq
Here the notation for the integrals indicates that the real-emission contribution involves $m+1$ final-state partons, while the other two terms have $m$-parton kinematics, and the phase spaces are restricted by the corresponding jet functions $J_n$ that define the physical quantity. The collinear subtraction term is explicitly given by the following expression
\beq
\bsp
\dsig{C}_{ab}(p_a,p_b;\muF{2}) &=
	-\frac{\as}{2\pi}S_\eps\sum_{c,d}\int_0^1 \rd z_a\,\int_0^1\rd z_b\,
	\dsig{B}_{cd}(z_a\,p_a,z_b\,p_b)
\\&\times
	\bigg\{\delta_{bd}\,\delta(1-z_b)
	\bigg[-\frac{1}{\eps}\bigg(\frac{\mu^2}{\muF{2}}\bigg)^\eps P^{ac}(z_a) 
	+ K^{ac}_{\rm{F.S.}}(z_a)\bigg]
\\&\quad+
	\delta_{ac}\,\delta(1-z_a)
	\bigg[-\frac{1}{\eps}\bigg(\frac{\mu^2}{\muF{2}}\bigg)^\eps P^{bd}(z_b) 
	+ K^{bd}_{\rm{F.S.}}(z_b)\bigg]\bigg\}\,.
\label{eq:dsigC}
\esp
\eeq
Above $S_\eps$ is the phase space factor due to the integral over the $(d-3)$-dimensional solid angle, which is included in the definition of the running coupling in the \msbar\ renormalization scheme\footnote{The \msbar\ renormalization scheme as often employed in the literature uses $S_\eps = (4\pi)^\eps \e^{-\eps\gamma_E}$. It is not difficult to check that the two definitions lead to the same expressions in a computation at NLO accuracy.}
\beq
S_\eps = \int\frac{\rd^{d-3}\Omega}{(2\pi)^{d-3}} 
	= \frac{(4\pi)^\eps}{\Gamma(1-\eps)}\,.
\label{eq:Sepsdef}
\eeq
The functions $P^{ab}(z)$ appearing in \eqn{eq:dsigC} are the {\em four-dimensional} Altarelli--Parisi probabilities given by
\bal
P^{qg}(x) &= \CF \frac{1+(1-x)^2}{x}\,,
\label{eq:Psplitqg}
\\[.5em]
P^{gq}(x) &= \TR\left[x^2+(1-x)^2\right]\,,
\label{eq:Psplitgq}
\\[.5em]
P^{qq}(x) &= \CF \left(\frac{1+x^2}{1-x}\right)_+
\label{eq:Psplitqq}
\eal
and
\beq
P^{gg}(x) = 2\CA\left[\left(\frac{1}{1-x}\right)_+ + \frac{1-x}{x} -1 + x(1-x)\right]
	+ \delta(1-x)\left(\frac{11}{6}\CA-\frac{2}{3}\Nf\TR\right)\,,
\label{eq:Psplitgg}
\eeq
where the `$+$'-distribution appearing above is defined, as usual, by its action on a generic test function $g(x)$
\beq
\int_0^1\rd x\,g(x)[f(x)]_+ \equiv \int_0^1\rd x\,[g(x)-g(1)]f(x)\,.
\label{eq:plusdist}
\eeq
The factorization scheme-dependent terms $K^{ab}_\FS$ are zero in the \msbar\ scheme. The parton densities are also scale/scheme dependent, so that this dependence cancels in the full hadronic cross section of \eqn{eq:xsec}.

%
%

\subsection{The subtraction procedure}
\label{ssec:subproc}

As is well known, the three terms of the NLO correction in \eqn{eq:tsigNLO} are all separately divergent, only their sum is finite for infrared-safe observables. The traditional approach to finding the finite NLO correction is to first continue all integrals in \eqn{eq:tsigNLO} to $d=4-2\eps$ dimensions, then regularize the real-emission contribution by subtracting a suitably defined approximate cross section, $\dsiga{R}{}_{ab}(p_a,p_b)$, such that
\begin{itemize}
	\item $\dsiga{R}{}_{ab}(p_a,p_b)$ matches the point-wise singularity structure of $\dsig{R}_{ab}(p_a,p_b)$ in the one-parton unresolved regions of phase space in $d$ dimensions,
	\item $\dsiga{R}{}_{ab}(p_a,p_b)$ can be integrated over the one-parton phase space of the unresolved parton analytically.
\end{itemize}
Then, to compute the NLO cross section we write
\beq
\bsp
\tsig{NLO}_{ab}(p_a,p_b;\muF{2}) &=
	\int_{m+1}\left[\dsig{R}_{ab}(p_a,p_b)J_{m+1}
	-\dsiga{R}{}_{ab}(p_a,p_b)J_m\right]_{\eps=0}
\\&+
	\int_m\left[\dsig{V}_{ab}(p_a,p_b)+\dsig{C}_{ab}(p_a,p_b;\muF{2})
	+\int_1\dsiga{R}{}_{ab}(p_a,p_b)\right]_{\eps=0}J_m\,.
\label{eq:subtract}
\esp
\eeq
Since the first integral on the right hand side of this equation is finite in four dimensions by construction, it follows from the Kinoshita--Lee--Nauenberg theorem that the combination of terms in the second integral is finite as well for any infrared-safe observable. 

The final result is that we have rewritten the NLO contribution as a sum of two finite integrals, both of which are integrable in four dimensions using standard numerical techniques.

Lastly, let us note that in \eqn{eq:subtract} the notation $\dsiga{R}{}_{ab}(p_a,p_b)J_m$ is symbolic in the sense that $\dsiga{R}{}_{ab}(p_a,p_b)$ is actually a sum of terms and the jet function multiplying each term depends on a different set of momenta. The precise definition is given in \sect{sec:approx_xsec}.


\section{Notation}
\label{sec:notation}

We adopt the colour- and spin-state notation of \refr{Catani:1996vz} with a {\em different normalization}.
In this notation the tree-level amplitude for the hadron-initiated scattering process involving the $n$ final-state momenta $\{p_1,\ldots,p_n\}$,
\beq
\cM_{n,ab}^{c_1,\ldots,c_n,c_a,c_b;s_1\ldots,s_n,s_a,s_b}(p_1,\ldots,p_n;p_a,p_b)\,,
\eeq
is an abstract vector in colour and spin space denoted by
\beq
\bsp
\ket{n,ab}{(0)}{(p_1,\ldots,p_n;p_a,p_b)} &\equiv
	(|c_1,\ldots,c_n,c_a,c_b\ra \otimes 
	|s_1\ldots,s_n,s_a,s_b\ra)
\\&\times
	\cM_{n,ab}^{c_1,\ldots,c_n,c_a,c_b;s_1\ldots,s_n,s_a,s_b}
	(p_1,\ldots,p_n;p_a,p_b)\,.
\label{eq:Mvector}
\esp
\eeq
Here the labels $a$ and $b$ refer to the initial-state partons. Notice that compared to the definition of \refr{Catani:1996vz} in their eq.\ (3.11), we {\em do not} include factors of $1/\sqrt{n_c(f_a)}$ for each initial-state parton $a$ of flavour $f_a$ carrying $n_c(f_a)$ colours. 

Colour interactions at QCD vertices are represented by associating colour charges $\bT_j$ with the emission of a gluon from each parton $j$. In the colour-state notation each vector $\ket{}{}{}$ is a colour singlet state, so colour conservation is simply
\beq
\sum_j\bT_j \ket{}{}{} = 0\,,
\label{eq:colourcons}
\eeq
where the sum runs over all external partons (initial- and final-state) of the vector $\ket{}{}{}$.

Owing to our normalization as given in \eqn{eq:Mvector} above, the two-parton colour-connected squared matrix element
\beq
\bra{n,ab}{(0)}{(p_1,\ldots,p_n;p_a,p_b)}
	\bT_i\bT_k\ket{n,ab}{(0)}{(p_1,\ldots,p_n;p_a,p_b)}
\eeq
{\em does not} include factors of $1/n_c(f_a)$ (compare with eq.\ (3.13) of \refr{Catani:1996vz}). The colour-charge algebra for the product $\bT_i\cdot\bT_k$ is
\beq
\bT_i\cdot\bT_k = \bT_k\cdot\bT_i\,,
\quad\mbox{if}\quad i\ne k\,;
\qquad\mbox{and}\qquad
\bT_i^2 = C_i\,,
\eeq
where $C_i$ is the quadratic Casimir operator in the representation of parton $i$. 
We have $\CF=\TR(\Nc^2-1)/\Nc = (\Nc^2-1)/(2\Nc)$ in the fundamental and $\CA = 2\TR\Nc = \Nc$ in the adjoint representation, i.e.\ we use the customary normalization $\TR=1/2$.

To indicate taking various kinematical limits, we use the formal operator notation introduced in \refr{Somogyi:2005xz}.


\section{Factorization in the collinear and soft limits}
\label{sec:fact}

We consider the tree-level squared matrix element of a generic hadron-initiated process with $m+1$ QCD partons in the final state $\SME{m+1,ab}{0}{\momab{}}$. Here $(\momab{})\equiv (p_1,\ldots,p_{m+1};p_a,p_b)$ denotes the set of $m+1$ final-state momenta and the momenta of the two incoming partons.

%
%

\subsection{The collinear limit}


\subsubsection{Final-final collinear}

Let us consider the limit when the momenta of the two final-state partons $i$ and $r$ become collinear. This limit is precisely defined by letting $\kT\to 0$ in the usual Sudakov parametrization of the momenta
\beq
p_i^\mu = z_i p_{ir}^\mu + \kT^\mu 
	- \frac{\kT^2}{z_i}\frac{n^\mu}{2p_{ir}\cdot n}\,,
\qquad
p_r^\mu = (1-z_i) p_{ir}^\mu - \kT^\mu 
	- \frac{\kT^2}{1-z_i}\frac{n^\mu}{2p_{ir}\cdot n}\,,
\label{eq:Sudakov}
\eeq
where $p_{ir}^\mu$ is the light-like ($p_{ir}^2=0$) momentum pointing in the collinear direction, while $n^\mu$ is an auxiliary light-like vector used to specify how the collinear direction is approached. The transverse momentum $\kT^\mu$ is orthogonal to both $p_{ir}^\mu$ and $n^\mu$ ($p_{ir}\cdot\kT=n\cdot\kT=0$).
In the small-$\kT$ limit (neglecting terms less singular than $1/\kT^2$), we find that the squared matrix element behaves as follows
\beq
\bsp
&\bC{ir}\SME{m+1,ab}{0}{\momab{}} = 
\\&\qquad =
	\pref\frac{1}{s_{ir}}\bra{m,ab}{(0)}{(\momab{m})}\hP_{f_if_r}(z_i,\kT;\eps)
	\ket{m,ab}{(0)}{(\momab{m})}\,.
\label{eq:CirM}
\esp
\eeq
Here $\hP_{f_if_r}(z_i,\kT;\eps)$ are the (tree-level) Altarelli--Parisi splitting kernels (that we choose to label by the flavours of the two daughter partons) and the $m$-parton matrix elements on the right hand side are obtained from $|\cM_{m+1,ab}^{(0)}(\momab{})\ra$ by removing partons $i$ and $r$ and replacing them with a single parton $ir$. This parton carries the quantum numbers of the pair $i+r$ in the collinear limit, that is its momentum is $p_{ir}^\mu$ and its other quantum numbers (colour, flavour) are determined according to the rule: gluon~+~anything gives gluon and quark~+~antiquark gives gluon.

The Altarelli--Parisi splitting kernels are matrices that act on the spin indices of the parton $ir$. Their explicit expressions are 
\bal
\la s|\hP_{qg}(z,\kT;\eps)|\bar{s}\ra &= \delta_{s\bar{s}}\,\CF
	\left[\frac{1+z^2}{1-z}-\eps(1-z)\right]\,,
\label{eq:PqgLOR}
\\[.5em]
\la s|\hP_{gq}(z,\kT;\eps)|\bar{s}\ra &= \delta_{s\bar{s}}\,\CF
	\left[\frac{1+(1-z)^2}{z}-\eps z\right]\,,
\label{eq:PgqLOR}
\\[.5em]
\la\mu|\hP_{q\qb}(z,\kT;\eps)|\nu\ra &= \TR
	\left[-g^{\mu\nu}+4z(1-z)\frac{\kT^\mu \kT^\nu}{\kT^2}\right]\,,
\label{eq:PqqLOR}
\\[.5em]
\la\mu|\hP_{gg}(z,\kT;\eps)|\nu\ra &= 2\CA
	\left[-g^{\mu\nu}\left(\frac{z}{1-z}+\frac{1-z}{z}\right)
	-2(1-\eps)z(1-z)\frac{\kT^\mu \kT^\nu}{\kT^2}\right]\,.
\label{eq:PggLOR}
\eal


\subsubsection{Initial-final collinear}

Now consider the case when the final-state momentum $p_r^\mu$ becomes collinear with the initial-state momentum $p_a^\mu$. Here the collinear limit is defined as follows
\beq
p_r^\mu = (1-x)p_a^\mu + \kT^\mu - \frac{\kT^2}{1-x}\frac{n^\mu}{2p_a\cdot n}\,,
\qquad 
p_{ar}^\mu = x p_a^\mu\,,
\label{eq:Sudakov_init}
\eeq
with $\kT\to 0$, and the corresponding splitting process $a\to ar + r$ involves a transition from the initial-state parton $a$ to the initial-state parton $ar$ together with the emission of the final-state parton $r$. The factorization formula appropriate in the collinear limit of \eqn{eq:Sudakov_init} reads
\beq
\bsp
&\bC{ar}\SME{m+1,ab}{0}{\momab{}} = 
\\&\qquad = 
	\pref\frac{1}{x}\frac{1}{s_{ar}}
	\bra{m,(ar)b}{(0)}{(\{p\}_m;p_{ar},p_b)}\hP_{f_{ar} f_r}(x,\kT;\eps)
	\ket{m,(ar)b}{(0)}{(\{p\}_m;p_{ar},p_b)}\,,
\label{eq:CarM}
\esp
\eeq
where the initial-state splitting kernels $\hP_{f_{ar}f_r}(x,\kT;\eps)$ are related to the time-like Altarelli--Parisi splitting kernels by
\beq
\hP_{f_{ar}f_r}(x,\kT;\eps) = 
	-(-1)^{F(f_a)+F(f_{ar})}
	\,x\,\hP_{f_a \bar{f}_r}(1/x,\kT;\eps)\,.
\label{eq:P_IF}
\eeq
Notice that similarly to the case of final-state splitting, we choose to label the initial-final splitting kernel by the flavours of the two daughter partons. In \eqn{eq:P_IF} above $F(q)=1$ and $F(g)=0$, which takes care of a factor of $(-1)$ if a fermionic line is crossed. Note that we do not need to include the customary factor of 
\beq
\frac{\omega(f_{ar})}{\omega(f_{a})} = 
	\frac{n_s(f_{ar})n_c(f_{ar})}{n_s(f_a)n_c(f_a)}
\eeq
in \eqn{eq:P_IF} that would account for the correct counting of the number of colours and spins of the incoming partons, because our matrix elements are defined without these factors. Nevertheless, for further reference, we record here that in conventional dimensional regularization
\beq
\omega(q)=n_s(q)n_c(q)=2\Nc\,,
\qquad\mbox{and}\qquad
\omega(g)=n_s(g)n_c(g)=2(1-\eps)(\Nc^2-1)\,.
\label{eq:omegansnc}
\eeq

The $m$-parton matrix element on the right hand side of \eqn{eq:CarM} is obtained from $|\cM_{m+1,ab}^{(0)}(\momab{})\ra$ by removing the final-state parton $r$ and replacing the initial-state parton $a$ with the parton $ar$. This parton carries the momentum $p_{ar}^\mu$.

%
%

\subsection{The soft limit}

Suppose that the momentum of the final state gluon $r$ becomes soft, i.e.\ we have $p_r^\mu =\lambda q^\mu$ with $q^\mu$ fixed and $\lambda\to 0$. In this limit the squared matrix element behaves as follows
\beq
\bsp
&\bS{r}\SME{m+1,ab}{0}{\momab{}} = 
\\&\qquad = 
	-\pref\sum_i\sum_{k\ne i}\frac{1}{2}\calS_{ik}(r)
	\bra{m,ab}{(0)}{(\momab{m})}\bT_i\bT_k\ket{m,ab}{(0)}{(\momab{m})}\,,
\label{eq:SrM}
\esp
\eeq
where
\beq
\calS_{ik}(r) = \frac{2s_{ik}}{s_{ir}s_{kr}}
\label{eq:Sikr}
\eeq
is the usual eikonal factor and the $m$-parton matrix elements on the right hand side are obtained from $|\cM_{m+1,ab}^{(0)}(\momab{})\ra$ by simply removing the soft gluon. The sums over $i$ and $k$ in \eqn{eq:SrM} above run over all initial and final state partons.

%
%

\subsection{Matching the limits}

It is well known that the collinear factorization formulae contain pieces that are singular in the soft limit, and conversely, the soft factorization formula includes collinear divergences. To avoid double subtraction in the regions of phase space where the collinear and soft limits overlap, the common soft-collinear contributions have to be identified and the overlaps disentangled. A general and efficient way to do this at higher perturbative orders was presented in \refr{Nagy:2007mn}. However at NLO the separation of limits is quite straightforward and here we simply follow the approach of \refr{Somogyi:2005xz} and compute the iterated limits explicitly.

First we consider the soft limits of the collinear factorization formulae. In the case of the final-final collinear formula, as $p_r^\mu\to 0$ we have $z_i\to 1$ and
\beq
\bS{r}\bC{ir}\SME{m+1,ab}{0}{\momab{}} = 
	\pref\frac{1}{s_{ir}}\frac{2}{1-z_i}\,\bT_i^2\,\SME{m,ab}{0}{\momab{m}}\,,
\label{eq:SrCirM}
\eeq
if $r$ is a gluon, while $\bS{r}\bC{ir}\SME{m+1,ab}{0}{\momab{}}=0$ if $r$ is a (anti)quark. The matrix element on the right hand side is obtained from the original $m+1$-parton matrix element by simply dropping parton $r$.
The $p_r^\mu\to 0$, $x\to 1$ soft limit of the initial-final collinear formula reads
\beq
\bS{r}\bC{ar}\SME{m+1,ab}{0}{\momab{}} = 
	\pref\frac{1}{s_{ar}}\frac{2}{1-x}\,\bT_a^2\,\SME{m,ab}{0}{\momab{m}}\,,
\label{eq:SrCarM}
\eeq
if $r$ is a gluon, and zero otherwise. Again the matrix element on the right hand side is obtained from the original $m+1$-parton matrix element by simply dropping parton $r$.

Next we compute the collinear limits of the soft factorization formula. In the case where the momenta of the final-state partons $i$ and $r$ become collinear, $p_i^\mu\to z_i p_{ir}^\mu$ and $p_r^\mu\to (1-z_i)p_{ir}^\mu$, we find
\beq
\bC{ir}\bS{r}\SME{m+1,ab}{0}{\momab{}} =  
	\pref\frac{1}{s_{ir}}\frac{2z_i}{1-z_i}\,\bT_i^2\,
	\SME{m,ab}{0}{\momab{m}}\,. 
\label{eq:CirSrMfin}
\eeq
When the momentum of the final-state parton $r$ becomes collinear with the momentum of the initial-state parton $a$, i.e.\ $p_r^\mu\to (1-x)p_a^\mu$, we obtain
\beq
\bC{ar}\bS{r}\SME{m+1,ab}{0}{\momab{}} =  
	\pref\frac{1}{s_{ar}}\frac{2}{1-x}\,\bT_a^2\,
	\SME{m,ab}{0}{\momab{m}}\,. 
\label{eq:CarSrMfin}
\eeq
The matrix elements on the right hand sides of \eqns{eq:CirSrMfin}{eq:CarSrMfin} are again obtained from the original $m+1$-parton matrix element by dropping parton $r$.

It is straightforward that the candidate subtraction term ($I$ and $F$ denote the set of initial- and final-state partons respectively)
\beq
\bsp
&\bA{1}\SME{m+1,ab}{0}{\momab{}} = 
\\&\qquad = 
	\Bigg[\sum_{i\in F}\sum_{\substack{r\in F \\ r\ne i}}\frac{1}{2}\bC{ir}
	+\sum_{a\in I}\sum_{r\in F}\bC{ar}
	+\sum_{r\in F}\Bigg(\bS{r}-\sum_{\substack{i\in F \\ i\ne r}}\bC{ir}\bS{r}
	-\sum_{a\in I}\bC{ar}\bS{r}\Bigg)\Bigg]
\\&\qquad\times
\SME{m+1,ab}{0}{\momab{}}\,
\label{eq:A1SME}
\esp
\eeq
counts each singly-unresolved limit precisely once and is free of double subtractions. This equation is an obvious generalization of eq.\ (4.1) of \refr{Somogyi:2006cz} to the case where coloured particles are allowed in the initial state. Nonetheless, \eqn{eq:A1SME} above cannot yet serve as a true subtraction counterterm because each term in the sum is only well-defined in a specific collinear and/or soft limit. To define a proper counterterm, we must first give unambiguous meaning to all expressions away from the limits as well. 


\section{NNLO-compatible subtraction terms}
\label{sec:subterms}

%
%

\subsection{Collinear subtractions}


\subsubsection{Final-final collinear}

We define the extension of the collinear factorization formula of \eqn{eq:CirM} as follows
\beq
\bsp
&\cC{ir}{FF}(\momab{}) = 
\\&\qquad = 
	\pref\frac{1}{s_{ir}}\bra{m,ab}{(0)}{(\momabt{m}{(ir)})}
	\hP_{f_if_r}(\zt{i},\kTt{i};\eps)
	\ket{m,ab}{(0)}{(\momabt{m}{(ir)})}
\\&\qquad\times
	(1-\alpha_{ir})^{2d_0-2(m-1)(1-\eps)}\Theta(\alpha_0-\alpha_{ir})\,,
\label{eq:Cir}
\esp
\eeq
where $\hP_{f_i f_r}$ are the Altarelli--Parisi kernels as given in \eqnss{eq:PqgLOR}{eq:PggLOR}. The momentum fractions $\zt{i}$ and $\zt{r}$ are
\beq
\zt{i} = \frac{y_{iQ}}{y_{(ir)Q}}\,,
\qquad\mbox{and}\qquad
\zt{r} = 1-\zt{i} = \frac{y_{rQ}}{y_{(ir)Q}}\,,
\label{eq:ztdef}
\eeq
where we define $y_{nQ} = s_{nQ}/Q^2 \equiv 2p_n\cdot Q/Q^2$ for $n=i,r$ and $y_{(ir)Q} = y_{iQ}+y_{rQ}$. The transverse momentum $\kTtm{i}{\mu}$ is
\beq
\kTtm{i}{\mu} = \zeta_{i,r}p_r^\mu - \zeta_{r,i} p_i^\mu + \zeta_{ir} \ti{p}_{ir}^\mu\,,
\qquad
\zeta_{i,r} = \zt{i}-\frac{y_{ir}}{\alpha_{ir}y_{(ir)Q}}\,,
\qquad
\zeta_{r,i} = \zt{r}-\frac{y_{ir}}{\alpha_{ir}y_{(ir)Q}}\,,
\label{eq:kTtdef}
\eeq
where $y_{ir} = s_{ir}/Q^2 \equiv 2p_i\cdot p_r/Q^2$, while the parent momentum $\ti{p}_{ir}^\mu$ and $\alpha_{ir}$ are defined below in \eqns{eq:pirtilde1}{eq:alphair} respectively. Above $Q$ is the total {\em partonic} incoming momentum in the {\em laboratory} center-of-mass frame, i.e.\ $Q^\mu=p_a^\mu+p_b^\mu \equiv (E_a+E_b,0,0,E_a-E_b)$. This choice of transverse momentum is exactly orthogonal to the parent momentum $(\kTt{i}\cdot \ti{p}_{ir}=0)$ and in the $p_i||p_r$ limit behaves as required $(\kTt{i}^2\to -z_i(1-z_i)s_{ir})$ for any $\zeta_{ir}$. Furthermore, the longitudinal component of $\kTt{i}$ proportional to $\zeta_{ir}$ vanishes by gauge-invariance when contracted with the matrix element, so in an NLO computation we can set $\zeta_{ir}=0$. However by choosing
\beq
\zeta_{ir} = \frac{y_{ir}}{\alpha_{ir}y_{\wti{ir}Q}}(\zt{r}-\zt{i})
\eeq
and using the Sudakov parametrization of \eqn{eq:Sudakov}, we find that $\kTtm{i}{\mu}\to 0$ in the $p_i||p_r$ limit (i.e.\ there is no residual `gauge term' proportional to $\ti{p}_{ir}^\mu$ in the limit). This is necessary if one wants to consider a generalization of the subtraction scheme to NNLO accuracy \cite{Somogyi:2006da}.

The matrix elements on the right hand side of \eqn{eq:Cir} are obtained from the original $m+1$-parton matrix element $|\cM_{m+1}^{(0)}({\momab{}})\ra$ by removing the partons $i$ and $r$ and replacing them by a single parton $ir$ as explained below \eqn{eq:CirM}. The momenta entering the $m$-parton matrix elements, $(\momabt{m}{(ir)}) \equiv (\ti{p}_1,\ldots,\ti{p}_{ir},\ldots,\ti{p}_{m+1};p_a,p_b)$ are defined as follows. First of all, it is convenient to leave the momenta of the incoming partons {\em unchanged}. Secondly, when defining the new final-state momenta, we find it useful to use a `democratic' mapping which treats the momenta of all (final-state) partons different from $i$ and $r$ identically. This choice is very convenient if one wants to extend the subtraction scheme to NNLO accuracy. Furthermore, in the context of an NLO computation, it leads to a smaller number of distinct phase space points at which subtraction terms have to be evaluated than the original `dipole' mapping of \Ref{Catani:1996vz}. We use the mapping introduced in \refr{Somogyi:2006cz}, where
\beq
\ti{p}_{ir}^\mu = \frac{1}{1-\alpha_{ir}}(p_i^\mu+p_r^\mu-\alpha_{ir}Q)\,,
\qquad
\ti{p}_n^\mu = \frac{1}{1-\alpha_{ir}}p_n^\mu\,,\qquad n\ne i,r,a,b\,,
\label{eq:pirtilde1}
\eeq
with
\beq
\alpha_{ir} = \frac{1}{2}\left[y_{(ir)Q}-\sqrt{y_{(ir)Q}^2-4y_{ir}}\;\right]\,.
\label{eq:alphair}
\eeq

Finally, note the two factors on the last line of \eqn{eq:Cir}. These are included to render the integrated counterterm $m$-independent (see \eqn{eq:IcCirFF} below) and to reduce the CPU time necessary for the numerical implementation \cite{Nagy:1998bb}. For further details, we refer the reader to appendix A of \refr{Somogyi:2008fc}.

~\\
\noindent{\bf Phase space factorization.}
The momentum mapping of \eqn{eq:pirtilde1} implements exact momentum conservation and leads to an exact factorization of the original $m+1$ parton phase space of total momentum $Q=p_a+p_b$ in the form
\beq
\PS{m+1}(\mom{};Q) = \PS{m}(\momt{(ir)}_m;Q)
	[\rd p_{1;m}^{(ir)}(p_r,\ti{p}_{ir};Q)]\,.
\label{eq:PSfactir}
\eeq
The factorized phase space measure can be written in several equivalent forms, here we choose the following \cite{Somogyi:2008fc,Aglietti:2008fe}
\beq
[\rd p_{1;m}^{(ir)}(p_r,\ti{p}_{ir};Q)] = 
	\rd \alpha_{ir}(1-\alpha_{ir})^{2(m-1)(1-\eps)-1} \frac{s_{\wti{ir}Q}}{2\pi}
	\PS{2}(p_i,p_r;p_{(ir)})\Theta(\alpha_{ir})\Theta(1-\alpha_{ir})\,,
\eeq
where $s_{\wti{ir}Q} = 2\ti{p}_{ir}\cdot Q$ and $p_{(ir)}^\mu = (1-\alpha_{ir})\ti{p}_{ir}^\mu + \alpha_{ir}Q^\mu$.

~\\
\noindent{\bf Integral of the subtraction term.}
Finally we compute the integral of the subtraction term over the unresolved phase space $[\rd p_{1;m}^{(ir)}(p_r,\ti{p}_{ir};Q)]$. Firstly, since $\kTt{i}^\mu$ as given by \eqn{eq:kTtdef} is orthogonal to $\ti{p}_{ir}^\mu$, the spin correlations generally present in \eqn{eq:Cir} vanish after azimuthal integration \cite{Catani:1996vz}. Thus when computing the integral of the subtraction term $\cC{ir}{FF}(\momab{})$ over the unresolved phase space, we can replace the Altarelli--Parisi splitting kernels $\hP_{f_i f_r}$ by their azimuthally averaged counterparts $P_{f_i f_r}$. Then we have
\beq
\bsp
&\int [\rd p_{1;m}^{(ir)}(p_r,\ti{p}_{ir};Q)] \cC{ir}{FF}(\momab{}) = 
\\&\qquad =
	\frac{\as}{2\pi}S_\eps \left(\frac{\mu^2}{Q^2}\right)^\eps
	\IcC{ir}{FF}(y_{\wti{ir}Q};\eps)\, \bT_{ir}^2
	\SME{m,ab}{0}{\momabt{m}{(ir)}}\,,
\esp
\eeq
where $y_{\wti{ir}Q} = 2\ti{p}_{ir}\cdot Q/Q^2$ and
\beq
\bsp
\IcC{ir}{FF}(y_{\wti{ir}Q};\eps) &= \frac{(4\pi)^2}{S_\eps}(Q^2)^\eps
\\&\times
	\int_0^{\alpha_0}\rd \alpha_{ir}\,(1-\alpha_{ir})^{2d_0-1} \frac{s_{\wti{ir}Q}}{2\pi}
	\int\PS{2}(p_i,p_r;p_{(ir)})\frac{1}{s_{ir}}P_{f_i f_r}(\zt{i},\eps)\frac{1}{\bT_{ir}^2}\,.
\label{eq:IcCirFF}
\esp
\eeq
As claimed, $\IcC{ir}{FF}(y_{\wti{ir}Q};\eps)$ is independent of $m$ and in order to lighten the notation throughout the paper, we do not explicitly indicate its dependence on $\alpha_0$ and $d_0$.

The integral appearing in \eqn{eq:IcCirFF} has been computed in \refrs{Somogyi:2008fc,Aglietti:2008fe} (see also \cite{Somogyi:2006cz} for the case of $\alpha_0=1$) and here we only recall the final results. The pole parts of the integral are independent of $\alpha_0$ and $d_0$ and we have
\bal
\IcC{qg}{FF}(x;\eps) &= \left[\frac{1}{\eps^2} + \frac{3}{2\eps} - \frac{2}{\eps}\ln(x)
	+ \Oe{0}\right]\,,	
	\qquad 
	\IcC{gq}{FF}(x;\eps) = \IcC{qg}{FF}(x;\eps)\,,
\\[.5em]
\IcC{q\qb}{FF}(x;\eps) & = \frac{\TR}{\CA}\left[-\frac{2}{3\eps} + \Oe{0}\right]\,,
\\[.5em]
\IcC{gg}{FF}(x;\eps) & = \left[\frac{2}{\eps^2} + \frac{11}{3\eps} - \frac{4}{\eps}\ln(x)
	+ \Oe{0}\right]\,.
\eal
We present the finite parts for the specific choice of $d_0=3-3\eps$ in \appx{app:IcCfin}.


\subsubsection{Initial-final collinear}

The extension of the collinear factorization formula of \eqn{eq:CarM} is written as
\beq
\bsp
&\cC{ar}{IF}(\momab{}) = \pref\frac{1}{\xt{a}}\frac{1}{s_{ar}}
\\&\qquad \times
	\bra{m,(ar)b}{(0)}{(\momabtt{m}{(ar)}{ar}{b})}
	\hP_{f_{ar}f_r}(\xt{a},\kTt{a};\eps)
	\ket{m,(ar)b}{(0)}{(\momabtt{m}{(ar)}{ar}{b})}
\\&\qquad\times
	\Theta(y'_0-y_{ar})\,,
\label{eq:Car}
\esp
\eeq
where $\hP_{f_{ar} f_r}$ are the initial-state Altarelli--Parisi kernels as given in \eqn{eq:P_IF}. The momentum fraction $\xt{a}$ is given by
\beq
\xt{a}=1-y_{rQ}\,,
\label{eq:xtdef}
\eeq
while the transverse momentum $\kTtm{a}{\mu}$ reads
\beq
\kTtm{a}{\mu} = p_r^\mu-y_{rQ} \, p_a^\mu-y_{ar} \, Q^\mu\,.
\label{eq:kTtadef}
\eeq
Again $Q$ is the total incoming partonic momentum in the laboratory center of mass frame, that is $Q^\mu=p_a^\mu+p_b^\mu$ and thus $Q^2=2p_a\cdot p_b = 2p_a\cdot Q$.
The choice of $\kTt{a}^\mu$ in \eqn{eq:kTtadef} is orthogonal to the parent momentum $\ti{p}_{ar}^\mu$ ($\kTt{a}\cdot \ti{p}_{ar} = 0$) to be defined in \eqn{eq:partilde} below and has the correct behaviour in the $p_a||p_r$ limit: $\kTtm{a}{2}\to -(1-x)s_{ar}$. Furthermore, using the appropriate Sudakov-parametrization, \eqn{eq:Sudakov_init}, we find that in the collinear limit $\kTtm{a}{\mu}\to 0$ (again there is no `gauge term' in the limit). This is important if the present NLO algorithm is to be included as part of an NNLO subtraction scheme.

Finally, the matrix elements on the right hand side of \eqn{eq:Car} are obtained from the original $m+1$-parton matrix element on the left by removing the final-state parton $r$ and replacing the initial-state parton $a$ with the parton $ar$ as discussed below \eqn{eq:omegansnc}. The $m$ momenta $(\momabtt{m}{(ar)}{ar}{b}) \equiv (\ti{p}_1,\ldots,\ti{p}_{m+1};\ti{p}_{ar},p_b)$ ($p_r$ is missing from the set) entering the factorized matrix elements are defined as follows. Firstly, it is convenient to leave the momentum of parton $b$ {\em unchanged}. Secondly, it is also convenient to define the momentum of parton $ar$, $\ti{p}_{ar}^\mu$ to be parallel with $p_a^\mu$. Thus we set
\beq
\ti{p}_{ar}^\mu = \xt{a} p_a^\mu\,,
\qquad
\ti{p}_n^\mu = \Lambda[Q-p_r,Q-(1-\xt{a})p_a]^\mu_\nu \, p_n^\nu\,,\qquad n\ne a,b,r\,,
\label{eq:partilde}
\eeq
where $\Lambda(K,\hat{K})^\mu_\nu$ is
\beq
\Lambda(K,\hat{K})^\mu_\nu = g^\mu_\nu
	-\frac{2(K+\hat{K})^\mu(K+\hat{K})_\nu}{(K+\hat{K})^2}
	+\frac{2K^\mu\hat{K}_\nu}{K^2} \,.
\label{eq:Lambda}
\eeq
This is the matrix of a (proper) Lorentz transformation, that maps $\hat{K}^\mu$ to $K^\mu$, whenever $K^2=\hat{K}^2$. This mapping treats all final-state momenta except $r$ identically.

The $\Theta$-function is included on the last line of \eqn{eq:Car} to reduce CPU time in a numerical implementation.

~\\
\noindent{\bf Phase space convolution.}
The analogue of the phase space {\em factorization} in \eqn{eq:PSfactir} is the following phase space {\em convolution}
\beq
\PS{m+1}(\mom{};p_a+p_b) = \int_0^1 \rd x\,\PS{m}(\momt{(ar)}_m;xp_a+p_b)
	[\rd p_{1;m}^{(ar)}(p_r,p_a,x)]\,,
\label{eq:PSconvar}
\eeq
where the one-particle phase space measure can be written as
\beq
\bsp
[\rd p_{1;m}^{(ar)}(p_r,p_a,x)] &= 
	\frac{S_\eps}{(4\pi)^2}(Q^2)^{1-\eps}
	\rd y_{ar}\,\rd \xt{a}\,
	\Theta(\xt{a})\Theta(1-\xt{a})\Theta(y_{ar})
	\Theta\left(1-\frac{y_{ar}}{1-\xt{a}}\right)
\\&\times
	(1-x)^{-2\eps}
	\left[\frac{y_{ar}}{1-\xt{a}}\left(1-\frac{y_{ar}}{1-\xt{a}}\right)\right]^{-\eps}
	\delta(x-\xt{a})\,.
\esp
\eeq
Above we have $y_{ar} = 2p_a\cdot p_r/Q^2$.

We note that the momentum mapping in \eqn{eq:partilde} and consequently the phase space convolution in \eqn{eq:PSconvar} coincides exactly with the mapping and convolution of section 5.5 of \Ref{Catani:1996vz}. Our $\xt{a}$ is exactly their $x_{i,ab}$, after the replacement $i \to r$ and similarly, their $\ti{v}_i$ is just our $y_{ar}$ after the same replacement.

~\\
\noindent{\bf Integral of the subtraction term.}
When computing the integral of $\cC{ar}{IF}(\momab{})$ over the one parton factorized phase space $[\rd p_{1;m}^{(ar)}(p_r,p_a,x)]$, we may replace the splitting functions $\hP_{f_{ar}f_r}$ with their azimuthally averaged counterparts $P_{f_{ar}f_r}$ since $\kTt{a}^\mu$ is orthogonal to $\ti{p}_{ar}^\mu$, and the spin correlations generally present in \eqn{eq:Car} vanish after integration. We find
\beq
\bsp
&\int[\rd p_{1;m}^{(ar)}(p_r,p_a,x)]\cC{ar}{IF}(\momab{}) = 
\\&\qquad =
	\frac{\as}{2\pi}S_\eps \left(\frac{\mu^2}{Q^2}\right)^\eps
	\frac{\omega(f_a)}{\omega(f_{ar})}\,
	\frac{1}{x}\IcC{a(ar)}{I}(x;\eps)\, \bT_{ar}^2
	\SME{m,(ar)b}{0}{\momabtt{m}{(ar)}{ar}{b}}\,,
\esp
\eeq
with
\beq
\bsp
\IcC{a(ar)}{I}(x;\eps) &= 
	\frac{\omega(f_{ar})}{\omega(f_a)}
	\int_0^{y'_0}\rd y_{ar}\int_0^1\rd \xt{a}\,
	\Theta\left(1-\frac{y_{ar}}{1-\xt{a}}\right)(1-x)^{-2\eps}
\\&\times
	\left[\frac{y_{ar}}{1-\xt{a}}\left(1-\frac{y_{ar}}{1-\xt{a}}\right)\right]^{-\eps}
	\delta(x-\xt{a})\frac{1}{s_{ar}} P_{f_{ar}f_r}(\xt{a};\eps)\frac{1}{\bT_{ar}^2}\,,
\esp
\label{eq:IcCaarI}
\eeq
where we have introduced factors of $\omega(f_a)$ and $\omega(f_{ar})$ into the definition of $\IcC{a(ar)}{I}(x;\eps)$ above for later convenience. As before, to lighten the notation, we do not explicitly indicate the dependence of $\IcC{a(ar)}{I}(x;\eps)$ on $y'_0$.

The integral which appears in \eqn{eq:IcCaarI} above has been computed in \refr{Catani:1996vz} for $y'_0=1$ and in \refr{Nagy:2003tz} for general $y'_0$, so here we only recall the final results. The pole part of the integral is independent of $y'_0$ and reads
\bal
\IcC{qg}{I}(x;\eps) &= -\frac{1}{\eps}\frac{\CF}{\CA}\left[\frac{1+(1-x)^2}{x}\right] 
	+ \Oe{0}\,,
\\[.5em]
\IcC{gq}{I}(x;\eps) &= -\frac{1}{\eps}\frac{\TR}{\CF}
	\left[x^2+(1-x)^2\right] + \Oe{0}\,,
\\[.5em]
\IcC{qq}{I}(x;\eps) &= -\frac{1}{\eps}
	\left(\frac{1+x^2}{1-x}\right)_+ 
	+ \delta(1-x)\left(\frac{1}{\eps^2} + \frac{3}{2\eps}\right) + \Oe{0}\,,
\\[.5em]
\IcC{gg}{IF}(x;\eps) &= -\frac{2}{\eps}\left[\left(\frac{1}{1-x}\right)_+ 
	+\frac{1-x}{x} - 1 + x(1-x)\right] + \delta(1-x)\frac{1}{\eps^2} + \Oe{0}\,,
\eal
where the `$+$'-distribution appearing above is defined in \eqn{eq:plusdist}. Note that for $y'_0=1$, these functions are in fact identical to the $\widetilde{{\cal V}}^{a,ai}(x;\eps)$ functions introduced in \refr{Catani:1996vz},  up to a factor of $\bT_{ar}^2$ (after the usual replacement of $i\to r$). Explicitly we have $\IcC{a(ar)}{I}(x;\eps;y'_0=1)\bT_{ar}^2 = \widetilde{{\cal V}}^{a,ar}(x;\eps)$.
The finite parts of these expressions are given in \appx{app:IcCfin}.

%
%

\subsection{Soft-type subtractions}
\label{sec:cS-type}

Next we define the extension of the soft (\eqn{eq:SrM}) and soft-collinear (\eqns{eq:CirSrMfin}{eq:CarSrMfin}) limits. We call these `soft-type' subtractions because they all use the same `soft' momentum mapping, see \eqn{eq:ptilde} below. We set
\bal
\cS{r}{}(\momab{}) &=
	-\pref
	\sum_{i}\sum_{k\ne i} \frac{1}{2}\calS_{ik}(r)
\nt\\&\times
	\bra{m,ab}{(0)}{(\momabt{m}{(r)})}\bT_i\bT_k\ket{m,ab}{(0)}{(\momabt{m}{(r)})}
\nt\\&\times
	(1-y_{rQ})^{d'_0-(m-1)(1-\eps)}\Theta(y_0-y_{rQ})P(y_{rQ})\,,
\label{eq:Sr}
\\[.5em]
\cC{ir}{FF}\cS{r}{}(\momab{}) &=
	\pref\frac{1}{s_{ir}}\frac{2\zt{i}}{1-\zt{i}}\,\bT_i^2\,	\SME{m,ab}{0}{\momabt{m}{(r)}}
\nt\\&\times
	(1-y_{rQ})^{d'_0-(m-1)(1-\eps)}\Theta(y_0-y_{rQ})P(y_{rQ})\,,
\label{eq:CirSr}
\\[.5em]
\cC{ar}{IF}\cS{r}{}(\momab{}) &=
	\pref\frac{1}{s_{ar}}\frac{2}{1-\xt{a}}\,\bT_a^2\,
	\SME{m,ab}{0}{\momabt{m}{(r)}}
\nt\\&\times
	(1-y_{rQ})^{d'_0-(m-1)(1-\eps)}\Theta(y_0-y_{rQ})P(y_{rQ})\,.
\label{eq:CarSr}
\eal
The eikonal factor, $\calS_{ik}(r)$, and the momentum fractions $\zt{i}$ and $\xt{a}$ which appear in the above equations were defined in eqs.\ (\ref{eq:Sikr}, \ref{eq:ztdef}) and (\ref{eq:xtdef}) respectively. (Also recall that the sums over $i$ and $k$ in \eqn{eq:Sr} run over all initial and final state partons.)
The matrix elements on the right hand sides are obtained form the original $m+1$-parton matrix element by simply dropping the parton $r$. The momenta $(\momabt{m}{(r)}) \equiv (\ti{p}_1,\ldots,\ti{p}_{m+1};p_a,p_b)$ ($p_r$ is missing) entering the $m$-parton matrix elements are defined as follows:
\beq
\ti{p}_n^\mu = \Lambda[Q,(Q-p_r)/\lambda_r]^\mu_\nu\,(p^\nu/\lambda_r)\,,
\qquad 
\lambda_r = \sqrt{1-y_{rQ}}\,,
\qquad
n\ne r,a,b\,,
\label{eq:ptilde}
\eeq
where $\Lambda$ is defined in \eqn{eq:Lambda}. Notice that the momenta of the incoming partons are left {\em unchanged}. This mapping treats all final-state momenta except $p_r$ in an identical fashion.

In \eqnss{eq:Sr}{eq:CarSr} above, we have included some additional factors. The first two of these serve the same purpose as in the collinear case, see \eqn{eq:Cir}. Firstly, they render the integrated counterterms $m$-independent (see eqs.\ (\ref{eq:IcSr}, \ref{eq:IcCSFFdef}, \ref{eq:IcCSIFdef}) below) and secondly they reduce the necessary CPU time in a numerical implementation. 

The final factor of $P(y_{rQ})$ is included to control the exact form of the {\em integrated} soft-type subtraction terms. A similar approach, albeit in a different context, has been used in \refr{Dittmaier:1999mb}. In the present case this is crucial for obtaining an insertion operator that obeys universal collinear factorization. In order that the terms in \eqnss{eq:Sr}{eq:CarSr} above reduce to their correct forms in the soft limit, we must have $P(y_{rQ})\to 1$ as $y_{rQ}\to 0$. Hence we set
\beq
P(y_{rQ}) = 1-\sum_{n=1}^{N} a_n\, y_{rQ}^n\,,
\label{eq:Py}
\eeq
where we shall find below that the minimal number of terms necessary is $N=2$ and we will also determine $a_1$ and $a_2$. The $a_n$ will be seen to depend on the parameters $y_0$ and $d'_0$ as well as $\eps$. In order to keep the notation streamlined, we do not indicate these dependences.

~\\
\noindent{\bf Phase space factorization.}
The momentum mapping defined in \eqn{eq:ptilde} above leads to an exact factorization of the original $m+1$-parton phase space of total momentum $Q=p_a+p_b$ in the form
\beq
\PS{m+1}(\mom{};Q) = \PS{m}(\momt{(r)}_m;Q)
	[\rd p_{1;m}^{(r)}(p_r;Q)]\,.
\label{eq:PSfactr}
\eeq
We choose to write the factorized phase space measure as in \refrs{Somogyi:2008fc,Aglietti:2008fe}
\beq
[\rd p_{1;m}^{(r)}(p_r;Q)] = 
	\rd y_{rQ}\,(1-y_{rQ})^{(m-1)(1-\eps)-1}\frac{Q^2}{2\pi}\PS{2}(p_r,K;Q)
	\Theta(y_{rQ})\Theta(1-y_{rQ})\,,
\label{eq:dp1mr}
\eeq
where the time-like momentum $K$ is massive with $K^2 = (1-y_{rQ})Q^2$.

~\\
\noindent{\bf Integral of the subtraction term.}
Consider first the integration of the soft counterterm $\cS{r}{}(\momab{})$ over the one-parton unresolved phase space $[\rd p_{1;m}^{(r)}(p_r;Q)]$. We find
\beq
\bsp
&\int[\rd p_{1;m}^{(r)}(p_r;Q)] \cS{r}{}(\momab{}) =
	\frac{\as}{2\pi}S_\eps\left(\frac{\mu^2}{Q^2}\right)^\eps
\\&\qquad\times
	\Bigg[
	\sum_{i\in F}\sum_{\substack{k\in F \\ k\ne i}}\IcS{ik}{FF}(\Y{i}{k};\eps)
	\bra{m,ab}{(0)}{(\momabt{m}{(r)})}\bT_i\bT_k\ket{m,ab}{(0)}{(\momabt{m}{(r)})}
\\&\qquad\qquad +
	2\sum_{i\in F}\IcS{ai}{IF}(Y_{a\ti{i},Q};\eps)
	\bra{m,ab}{(0)}{(\momabt{m}{(r)})}\bT_a\bT_i\ket{m,ab}{(0)}{(\momabt{m}{(r)})}
\\&\qquad\qquad +
	2\sum_{i\in F}\IcS{bi}{IF}(Y_{b\ti{i},Q};\eps)
	\bra{m,ab}{(0)}{(\momabt{m}{(r)})}\bT_b\bT_i\ket{m,ab}{(0)}{(\momabt{m}{(r)})}
\\&\qquad\qquad +
	2\,\IcS{ab}{II}(\eps)
	\bra{m,ab}{(0)}{(\momabt{m}{(r)})}\bT_a\bT_b\ket{m,ab}{(0)}{(\momabt{m}{(r)})}
\Bigg]\,,
\esp
\label{eq:IcSfull}
\eeq
where $a$ and $b$ are the labels of the two incoming partons and
\beq
\IcS{ik}{J_i J_k}(\Y{i}{k};\eps) = -\frac{(4\pi)^2}{S_\eps}(Q^2)^\eps
	\int_0^{y_0}\rd y_{rQ}\,(1-y_{rQ})^{d'_0-1}P(y_{rQ})\frac{Q^2}{2\pi}
	\int\PS{2}(p_r,K;Q) \frac{1}{2}\calS_{ik}(r)\,.
\label{eq:IcSr}
\eeq
Here $J_i,J_k = I,F$ denote whether $i$ and $k$ are initial- or final-state partons. Note that $\IcS{ik}{J_i J_k}(\Y{i}{k};\eps)$ is symmetric in $i$ and $k$, so the ordering of indices does not matter. In particular $\IcS{ki}{FI}(Y;\eps) = \IcS{ik}{IF}(Y;\eps)$. As promised, $\IcS{ik}{J_i J_k}(\Y{i}{k};\eps)$ is independent of $m$ and depends on $i$ and $k$ only through the combination
\beq
\Y{i}{k}=\frac{y_{\ti{i}\ti{k}}}{y_{\ti{i}Q}y_{\ti{k}Q}}\,.
\label{eq:YikQdef}
\eeq 
Notice that $\ti{p}_a = p_a$, $\ti{p}_b = p_b$ (the soft mapping of \eqn{eq:ptilde} leaves the incoming momenta unchanged), so $y_{\ti{a}Q}=y_{\ti{b}Q}=1$ and thus $\Y{a}{b}=1$. Then in \eqn{eq:IcSfull}, $\IcS{ab}{II}$ only depends on $\eps$, as indicated. As usual, we do not show the dependence on $y_0$ and $d'_0$ to lighten the notation. 

Next, consider the integral of the soft-collinear subtraction terms. We have
\beq
\int[\rd p_{1;m}^{(r)}(p_r;Q)] \cC{ir}{FF}\cS{r}{}(\momab{}) =
	\frac{\as}{2\pi}S_\eps\left(\frac{\mu^2}{Q^2}\right)^\eps \IcCS{FF}(\eps)\,
	\bT_i^2 \SME{m,ab}{0}{\momabt{m}{(r)}}
\label{eq:IntCSFF}
\eeq
and
\beq
\int[\rd p_{1;m}^{(r)}(p_r;Q)] \cC{ar}{IF}\cS{r}{}(\momab{}) =
	\frac{\as}{2\pi}S_\eps\left(\frac{\mu^2}{Q^2}\right)^\eps \IcCS{IF}(\eps)\,
	\bT_a^2 \SME{m,ab}{0}{\momabt{m}{(r)}}
\label{eq:IntCSIF}
\eeq
with
\beq
\IcCS{FF}(\eps) = \frac{(4\pi)^2}{S_\eps}(Q^2)^\eps
	\int_0^{y_0}\rd y_{rQ}\,(1-y_{rQ})^{d'_0-1}P(y_{rQ})\frac{Q^2}{2\pi}
	\int\PS{2}(p_r,K;Q) \frac{1}{s_{ir}}\frac{2\zt{i}}{1-\zt{i}}\,,
\label{eq:IcCSFFdef}
\eeq
and
\beq
\IcCS{IF}(\eps) = \frac{(4\pi)^2}{S_\eps}(Q^2)^\eps
	\int_0^{y_0}\rd y_{rQ}\,(1-y_{rQ})^{d'_0-1}P(y_{rQ})\frac{Q^2}{2\pi}
	\int\PS{2}(p_r,K;Q) \frac{1}{s_{ar}}\frac{2}{1-\xt{i}}\,.
\label{eq:IcCSIFdef}
\eeq
As before, $\IcCS{FF}(\eps)$ and $\IcCS{IF}(\eps)$ are independent of $m$ but do depend on $y_0$ and $d'_0$. The dependence on the later two parameters however is suppressed in the notation.

The soft integral in \eqn{eq:IcSr} for $i$ and $k$ {\em both} final-state partons,
as well as the soft-collinear integral in \eqn{eq:IcCSFFdef} above have been evaluated in \refrs{Somogyi:2008fc,Aglietti:2008fe}, with $P(y_{rQ})\equiv 1$ (see also \cite{Somogyi:2006cz} for the case of $y_0=1$). Computing the soft integral for the cases when either one or both of $i$ and $k$ are initial-state momenta and the soft-collinear integral in \eqn{eq:IcCSIFdef} above, with $P(y_{rQ})$ as given in \eqn{eq:Py}, is straightforward with the same techniques. We nevertheless present the details here, as they are needed to derive the $a_n$ that enter $P(y_{rQ})$ in \eqn{eq:Py}.

First, in order to write the factorized phase space measure in \eqn{eq:dp1mr} explicitly, we choose a specific Lorentz-frame. The choice of a convenient frame is different for $\IcS{ik}{FF}(\Y{i}{k};\eps)$, $\IcS{ai}{IF}(Y_{a\ti{i},Q};\eps)$ and $\IcS{ab}{II}(\eps)$ as well as for $\IcCS{FF}(\eps)$ and $\IcCS{IF}(\eps)$. In particular, in each frame we have
\beq
Q^\mu = \sqrt{s}(1,\ldots)\,,
\qquad\mbox{and}\qquad
p_r^\mu = 
E_r(1,\mbox{..`angles'..},\sin\vartheta\sin\varphi,\sin\vartheta\cos\varphi,\cos\vartheta)\,,
\label{eq:frame_all}
\eeq
where the dots stand for vanishing components, while the notation ..`angles'.. denotes the dependence of $p_r^\mu$ on the $d-3$ angular variables that can be trivially integrated. Then, for computing the various integrals, we set
\bal
\IcS{ik}{FF}(\Y{i}{k};\eps)\, , \,\IcCS{FF}(\eps) &:
	\quad
	\ti{p}_i^\mu = \ti{E}_i(1,\ldots,1)\,,
&
	\ti{p}_k^\mu& = \ti{E}_k(1,\ldots,\sin\chi_{\ti{i}\ti{k}},\cos\chi_{\ti{i}\ti{k}})\,,
\label{eq:frame_ik}
\\
\IcS{ai}{IF}(Y_{a\ti{i},Q};\eps)\, , \,\IcCS{IF}(\eps) &:
	\quad
	p_a^\mu = \frac{\sqrt{s}}{2}(1,\ldots,1)\,,	
&
	\ti{p}_i^\mu& = \ti{E}_i(1,\ldots,\sin\chi_{a\ti{i}},\cos\chi_{a\ti{i}})\,,
\label{eq:frame_ai}
\\
\IcS{ab}{II}(\eps) &:
	\quad
	p_a^\mu = \frac{\sqrt{s}}{2}(1,\ldots,1)\,,
&
	p_b^\mu& = \frac{\sqrt{s}}{2}(1,\ldots,-1)\,.
\label{eq:frame_ab}
\eal
Now in terms of the scaled energy-like variable
\beq
\eps_r = \frac{2p_r\cdot Q}{Q^2} = \frac{2E_r}{\sqrt{s}}
\label{eq:yrQ}
\eeq
and the angular variables $\vartheta$ and $\varphi$ the two-particle
phase space reads
\beq
\bsp
\PS{2}(p_{r},K;Q) &=
\frac{(Q^{2})^{-\eps}}{16\pi^{2}} S_\eps
\frac{\Gamma^{2}(1-\eps)}{\Gamma(1-2\eps)}
\,\rd\eps_r\,\eps_r^{1-2\eps}\delta(y-\eps_r)\\
&\times
\rd(\cos\vartheta)\,\rd(\cos\varphi)
(\sin\vartheta)^{-2\eps}(\sin\varphi)^{-1-2\eps}
\,,
\label{eq:PS2Sr}
\esp
\eeq
for all three choices of frame in \eqnss{eq:frame_ik}{eq:frame_ab}. In \eqn{eq:PS2Sr}, the limits of integration on $\eps_r$ are $0$ and $1$, while the cosines of both angles run from $-1$ to $+1$.

To write the integrands in these variables, we observe that the precise definitions of $\ti{p}_i$ and $\ti{p}_k$ as given in \eqn{eq:ptilde} above imply 
\beq
s_{ik} = (1-\eps_r) s_{\ti{i}\ti{k}}\,,
\qquad
s_{ir} = s_{\ti{i}r}\,,
\qquad
s_{kr} = s_{\ti{k}r}\,,
\qquad
s_{iQ} = (1-\eps_r) s_{\ti{i}Q} + s_{\ti{i}r}\,,
\label{eq:siktilde}
\eeq
and
\beq
s_{ai} = \sqrt{1-\eps_r}\,s_{a\ti{i}}
	+\frac{1}{1+\sqrt{1-\eps_r}}\frac{s_{aQ}s_{\ti{i}r}}{Q^2}
	-\frac{\sqrt{1-\eps_r}}{1+\sqrt{1-\eps_r}}\frac{s_{ar}s_{\ti{i}Q}}{Q^2}
	-\frac{1}{(1+\sqrt{1-\eps_r})^2}\frac{s_{ar}s_{\ti{i}r}}{Q^2}\,.
\eeq
Then in the Lorentz-frame of \eqn{eq:frame_ik} we find
\beq
\frac{s_{ik}}{s_{ir}s_{kr}} =
	(1-\eps_r) \frac{s_{\ti{i}\ti{k}}}{s_{\ti{i}r}s_{\ti{k}r}} =
	\frac{4 \Y{i}{k}}{Q^2}\frac{(1-\eps_r)}{\eps_r^2}
	\frac{1}{(1-\cos\vartheta)(1-\cos\chi_{\ti{i}\ti{k}}\cos\vartheta-\sin\chi_{\ti{i}\ti{k}}\sin\vartheta\cos\varphi)}
\,,
\label{eq:Sikrtilde}
\eeq
and
\beq
\frac{1}{s_{ir}}\frac{\zt{i}}{1-\zt{i}} =
	\frac{1}{s_{\ti{i}r}}\frac{(1-\eps_r) s_{\ti{i}Q}+s_{\ti{i}r}}{s_{rQ}} =
	\frac{1}{Q^2}\frac{1}{\eps_r}
	\left[1+\frac{2(1-\eps_r)}{\eps_r(1-\cos\vartheta)}\right]\,.
\label{eq:csFFtilde}
\eeq
Similarly, in the frame of \eqn{eq:frame_ai} we obtain
\beq
\bsp
\frac{s_{ai}}{s_{ar}s_{ir}} &=
	\sqrt{1-\eps_r}\frac{s_{a\ti{i}}}{s_{ar}s_{\ti{i}r}}
	+\frac{1}{1+\sqrt{1-\eps_r}}\frac{s_{aQ}}{s_{ar}Q^2}
	-\frac{\sqrt{1-\eps_r}}{1+\sqrt{1-\eps_r}}\frac{s_{\ti{i}Q}}{s_{\ti{i}r}Q^2}
	-\frac{1}{(1+\sqrt{1-\eps_r})^2}\frac{1}{Q^2}\,,
\\&=
	\frac{4 Y_{a\ti{i},Q}}{Q^2}\frac{\sqrt{1-\eps_r}}{\eps_r^2}
	\frac{1}{(1-\cos\vartheta)(1-\cos\chi_{a\ti{i}}\cos\vartheta-\sin\chi_{a\ti{i}}\sin\vartheta\cos\varphi)}
\\&\quad+
	\frac{2}{\eps_r(1+\sqrt{1-\eps_r})Q^2}
	\left(\frac{1}{1-\cos\vartheta}
	-
	\frac{\sqrt{1-\eps_r}}{1-\cos\chi_{a\ti{i}}\cos\vartheta-\sin\chi_{a\ti{i}}\sin\vartheta\cos\varphi}\right)
\\&\quad-
	\frac{1}{(1+\sqrt{1-\eps_r})^2Q^2}\,,
\esp
\label{eq:Sairtilde}
\eeq
and
\beq
\frac{1}{s_{ar}}\frac{1}{1-\xt{a}} = \frac{1}{s_{ar}}\frac{1}{y_{rQ}} =
	\frac{2}{Q^2}\frac{1}{\eps_r^2}\frac{1}{1-\cos\vartheta}\,.
\label{eq:csIFtilde}
\eeq
Finally, in the frame of \eqn{eq:frame_ab} we have simply
\beq
\frac{s_{ab}}{s_{ar}s_{br}} = \frac{4}{Q^2}\frac{1}{\eps_r^2}
	\frac{1}{1-\cos^2\vartheta}\,.
\label{eq:Sabrtilde}
\eeq
The $\Y{i}{k}$ and $Y_{a\ti{i},Q}$ that appear in \eqns{eq:Sikrtilde}{eq:Sairtilde} above are defined in \eqn{eq:YikQdef}.

Now we are ready to compute the integrals in eqs.\ (\ref{eq:IcSr}, \ref{eq:IcCSFFdef}) and (\ref{eq:IcCSIFdef}). We start by computing $\IcS{ik}{FF}(\Y{i}{k};\eps)$ and $\IcCS{FF}(\eps)$. Using eqs.\ (\ref{eq:PS2Sr}, \ref{eq:Sikrtilde}) and (\ref{eq:csFFtilde}) we obtain
\beq
\bsp
\IcS{ik}{FF}(\Y{i}{k};\eps) &= 
	-4\Y{i}{k}\frac{\Gamma^2(1-\eps)}{2\pi\Gamma(1-2\eps)}\Omega^{(1,1)}(\cos\chi_{\ti{i}\ti{k}})
\\&\times
	\Big[B_{y_0}(-2\eps,d'_0+1)-\sum_{n=1}^N a_n\,B_{y_0}(n-2\eps,d'_0+1)\Big]
\esp
\label{eq:IcSikFF}
\eeq
and
\beq
\bsp
\IcCS{FF}(\eps) &=
	2B(-\eps,1-\eps)\Big[B_{y_0}(-2\eps,d'_0)-\sum_{n=1}^N a_n\,B_{y_0}(n-2\eps,d'_0)\Big]
\\&-
	2B(-\eps,2-\eps)\Big[B_{y_0}(1-2\eps,d'_0)-\sum_{n=1}^N a_n\,B_{y_0}(n+1-2\eps,d'_0)\Big]\,.
\esp
\label{eq:IcCSFF}
\eeq
In these eqs.\ $B_{y_0}(\alpha,\beta)$ is the incomplete beta function
\beq
B_{y_0}(\alpha,\beta) = \int_0^{y_0} \rd y\,y^{\alpha-1}(1-y)^{\beta-1}\,,
\label{eq:By0}
\eeq
while $\Omega^{(i,k)}(\cos\chi)$ in \eqn{eq:IcSikFF} above denotes the angular integral
\beq
\bsp
\Omega^{(i,k)}(\cos\chi) &=
	\int_{-1}^1\!\rd(\cos\vartheta)\;(\sin\vartheta)^{-2\eps}
	\int_{-1}^1\!\rd(\cos\varphi)\;(\sin\varphi)^{-1-2\eps}
\\&\times
	(1-\cos\vartheta)^{-i}
	(1-\cos\chi\cos\vartheta-\sin\chi\sin\vartheta\cos\varphi)^{-k}\,.
\label{eq:Omegaik}
\esp
\eeq
This was computed in \refr{vanNeerven:1985xr} and we have (note the different normalization of the angular integral here as compared to \cite{vanNeerven:1985xr})
\beq
\Omega^{(i,k)}(\cos\chi) =
	2^{1-i-k}\,\pi
	\frac{\Gamma(1-2\eps)}{\Gamma^2(1-\eps)}B(1-i-\eps,1-k-\eps)\,
	{}_2F_1\left(i,k,1-\eps,\frac{1+\cos\chi}{2}\right)\,.
\label{eq:Omegaikres}
\eeq
Furthermore, from \eqn{eq:frame_ik} it is easy to see that
\beq
\cos\chi_{\ti{i}\ti{k}} = 1-2\,\Y{i}{k}\,.
\label{eq:coschi}
\eeq
Next, consider $\IcS{ai}{IF}(Y_{a\ti{i},Q};\eps)$ and $\IcCS{IF}(\eps)$. Using eqs.\ (\ref{eq:PS2Sr}, \ref{eq:Sairtilde}) and (\ref{eq:csIFtilde}) we find
\beq
\bsp
\IcS{ai}{IF}(Y_{a\ti{i},Q};\eps) &= 
	-4Y_{a\ti{i},Q}\frac{\Gamma^2(1-\eps)}{2\pi\Gamma(1-2\eps)}\Omega^{(1,1)}(\cos\chi_{a\ti{i}})
\\&\times
	\Big[B_{y_0}(-2\eps,d'_0+\shalf)-\sum_{n=1}^N a_n\,B_{y_0}(n-2\eps,d'_0+\shalf)\Big]
\\&-
	\Big[B(-\eps,1-\eps)-B(1-\eps,1-\eps)\Big]
\\&\times
	\bigg\{B_{y_0}(-2\eps,d'_0)+B_{y_0}(-2\eps,d'_0+1)-2B_{y_0}(-2\eps,d'_0+\shalf)
\\&\qquad-
	\sum_{n=1}^{N}a_n\,\Big[B_{y_0}(n-2\eps,d'_0)+B_{y_0}(n-2\eps,d'_0+1)-2B_{y_0}(n-2\eps,d'_0+\shalf)\Big]\bigg\}
\esp
\label{eq:IcSaiIF}
\eeq
with $\cos\chi_{a\ti{i}} = 1-2Y_{a\ti{i},Q}$, and
\beq
\IcCS{IF}(\eps) =
	2B(-\eps,1-\eps)\Big[B_{y_0}(-2\eps,d'_0)-\sum_{n=1}^N a_n\,B_{y_0}(n-2\eps,d'_0)\Big]\,.
\label{eq:IcCSIF}
\eeq
Finally, using \eqns{eq:PS2Sr}{eq:Sabrtilde}, for $\IcS{ab}{II}(\eps)$ we obtain
\beq
\IcS{ab}{II}(\eps) = 
	-B(-\eps,-\eps)\Big[B_{y_0}(-2\eps,d'_0)-\sum_{n=1}^N a_n\,B_{y_0}(n-2\eps,d'_0)\Big]\,.
\label{eq:IcSabII}
\eeq

~\\
\noindent{\bf The choice of $\bom{a_n}$.}
Before presenting the $\eps$-expansion of the soft-type integrals, let us consider how the $a_n$s appearing in \eqn{eq:Py} are fixed. The crucial observation, as discussed in the Introduction, is that we must obtain an integrated approximate cross section which has a universal collinear limit. However, the colour-connected pieces of the integrated approximate cross section can ruin universal collinear factorization as explained below \eqn{eq:CirTkTiTkTr}. Now as \eqn{eq:IcSfull} shows, the integrated soft terms $\IcS{ik}{J_i J_k}$ will enter the integrated approximate cross section multiplying colour-connected matrix elements (see \eqn{eq:intAiii_1}). Thus, in order to guarantee that the integrated approximate cross section does have a universal collinear limit, we need to insist that
\beq
\bC{jl}\IcS{pj}{J_p J_j} = \bC{jl}\IcS{pl}{J_p J_l}\,,
\label{eq:condsALL}
\eeq
for {\em any} $j,l$ and $p$. This condition implies certain constraints on the $a_n$ that appear in the function $P(y_{rQ})$ in \eqn{eq:Py}.

To understand these constraints, consider first the case when $j,l$ and $p$ are all final-state partons. Then to satisfy \eqn{eq:condsALL} we need the following equality to hold:
\beq
\bC{\ti{j}\ti{l}}\IcS{pj}{FF}(\Y{p}{j};\eps) = \bC{\ti{j}\ti{l}}\IcS{pl}{FF}(\Y{p}{l};\eps)\,.
\label{eq:condFF}
\eeq
As is clear from \eqn{eq:YikQdef}, the variable $Y_{ik,Q}$ is homogeneous in the momenta $p_i^\mu$ and $p_k^\mu$, thus in the collinear limit, when $\ti{p}_j^\mu\to z \ti{p}_{jl}^\mu$ and $\ti{p}_l^\mu\to (1-z) \ti{p}_{jl}^\mu$, we simply find
\beq
\bC{\ti{j}\ti{l}}\Y{p}{j} = \bC{\ti{j}\ti{l}}\Y{p}{l} = Y_{\ti{p}\wti{jl},Q}\,,
\label{eq:CjlY}
\eeq
and \eqn{eq:condFF} is trivially satisfied. The same argument shows that if now $p$ is an initial-state parton,
\beq
\bC{\ti{j}\ti{l}}\IcS{pj}{IF}(Y_{p\ti{j},Q};\eps) = \bC{\ti{j}\ti{l}}\IcS{pl}{IF}(Y_{p\ti{l},Q};\eps)
\label{eq:condIF}
\eeq
also holds.

Next, let $p$ and e.g.\ $j$ be final-state partons, while $l$ is an initial-state one. By \eqn{eq:condsALL} we require
\beq
\bC{\ti{j}l}\IcS{pj}{FF}(\Y{p}{j};\eps) = \bC{\ti{j}l}\IcS{pl}{FI}(Y_{\ti{p}l,Q};\eps)
\equiv
\bC{\ti{j}l}\IcS{lp}{IF}(Y_{l\ti{p},Q};\eps)\,.
\label{eq:condFFIF}
\eeq
Of course \eqn{eq:CjlY} is again satisfied (recall that if $l$ is an initial-state parton then $\ti{p}_l = p_l$), but it is obvious from \eqns{eq:IcSikFF}{eq:IcSaiIF} that $\IcS{pj}{FF}(Y;\eps)$ and $\IcS{lp}{IF}(Y;\eps)$ are in general {\em different} functions of the variables $Y$ and $\eps$. We have included the $P(y_{rQ})$ function in the definition of the soft subtraction term, \eqn{eq:Sr}, precisely to have the freedom to enforce \eqn{eq:condFFIF} above. In particular, we if require that $\IcS{pj}{FF}(Y;\eps)$ and $\IcS{lp}{IF}(Y;\eps)$ are equal, then by virtue of \eqn{eq:CjlY}, \eqn{eq:condFFIF} holds. We find that $\IcS{pj}{FF}(Y;\eps) = \IcS{lp}{IF}(Y;\eps)$ is satisfied if we demand that
\beq
\bsp
&B_{y_0}(-2\eps,d'_0+1)-\sum_{n=1}^N a_n\,B_{y_0}(n-2\eps,d'_0+1) =
\\&\qquad=
B_{y_0}(-2\eps,d'_0+\shalf)-\sum_{n=1}^N a_n\,B_{y_0}(n-2\eps,d'_0+\shalf)\,,
\label{eq:betacond1}
\esp
\eeq
and
\beq
\bsp
0 &= B_{y_0}(-2\eps,d'_0)+B_{y_0}(-2\eps,d'_0+1)-2B_{y_0}(-2\eps,d'_0+\shalf)
\\&-
	\sum_{n=1}^{N}a_n\,\Big[B_{y_0}(n-2\eps,d'_0)+B_{y_0}(n-2\eps,d'_0+1)-2B_{y_0}(n-2\eps,d'_0+\shalf)\Big]\,.
\label{eq:betacond2}
\esp
\eeq

Finally, if $p$ and $l$ are the two initial-state partons while $j$ is a final-state one, \eqn{eq:condsALL} implies that we need to have
\beq
\bC{\ti{j}l}\IcS{pj}{IF}(Y_{p\ti{j},Q};\eps) = \bC{\ti{j}l}\IcS{pl}{II}(\eps)\,.
\label{eq:condII}
\eeq
In this limit we have $\ti{p}_j^\mu\to (1-x)p_l^\mu$, so we find
\beq
\bC{\ti{j}l}Y_{p\ti{j},Q} = Y_{pl,Q}\equiv 1\,,
\eeq
because $p$ and $l$ are the two initial-state partons (see the discussion below \eqn{eq:YikQdef}). But as can be seen from \eqns{eq:IcSaiIF}{eq:IcSabII}, $\IcS{pj}{IF}(1;\eps)$ and $\IcS{pl}{II}(\eps)$ are generally {\em different} functions of $\eps$. Demanding their equality given \eqn{eq:betacond2} implies
\beq
\bsp
&B_{y_0}(-2\eps,d'_0+\shalf)-\sum_{n=1}^N a_n\,B_{y_0}(n-2\eps,d'_0+\shalf) =
\\&\qquad=
B_{y_0}(-2\eps,d'_0)-\sum_{n=1}^N a_n\,B_{y_0}(n-2\eps,d'_0)\,.
\label{eq:betacond3}
\esp
\eeq
We have used that 
\beq
4\frac{\Gamma^2(1-\eps)}{2\pi\Gamma(1-2\eps)}\Omega^{(1,1)}(-1) = B(-\eps,-\eps)\,.
\eeq

Thus, we find that if we choose the $a_n$ such that \eqns{eq:betacond1}{eq:betacond2} as well as \eqn{eq:betacond3} are satisfied, then
\beq
\IcS{ik}{FF}(Y;\eps) = \IcS{ai}{IF}(Y;\eps)
\eeq
and
\beq
\IcS{ai}{IF}(1;\eps) = \IcS{ab}{II}(\eps)\,,
\eeq
so the collinear limit of the integrated soft counterterm is universal.

To solve for the $a_n$, note first that eqs.\ (\ref{eq:betacond1}, \ref{eq:betacond2}) and (\ref{eq:betacond3}) are not independent. In fact, \eqns{eq:betacond1}{eq:betacond3} clearly imply \eqn{eq:betacond2}. Therefore we only have two independent equations and thus the minimal number of $a_n$s is $N=2$. Here we limit ourselves to this minimal solution and use \eqn{eq:betacond3} to get
\beq
\bsp
B_{y_0}(-2\eps,d'_0+\shalf) - B_{y_0}(-2\eps,d'_0) &=
	\Big[B_{y_0}(1-2\eps,d'_0+\shalf) - B_{y_0}(1-2\eps,d'_0)\Big]a_1
\\&\,+
	\Big[B_{y_0}(2-2\eps,d'_0+\shalf) - B_{y_0}(2-2\eps,d'_0)\Big]a_2
\label{eq:C1}
\esp
\eeq
and take the difference of \eqns{eq:betacond3}{eq:betacond1} to obtain
\beq
B_{y_0}(1-2\eps,d'_0) = B_{y_0}(2-2\eps,d'_0)a_1 + B_{y_0}(3-2\eps,d'_0)a_2
\label{eq:C2}
\eeq
as the two independent equations determining $a_1$ and $a_2$. In deriving \eqn{eq:C2} we use the identity
\beq
B_{y_0}(\alpha,\beta) - B_{y_0}(\alpha,\beta+1) = B_{y_0}(\alpha+1,\beta)\,,
\eeq
which is easy to show starting from the integral representation of the incomplete beta function as given in \eqn{eq:By0}. It is straightforward to solve \eqns{eq:C1}{eq:C2} for $a_1$ and $a_2$, and we give the solution in \appx{app:IcSfin}. As remarked before, $a_1$ and $a_2$ depend on $y_0$, $d'_0$ and $\eps$. It is important to point out that for $d'_0 = D'_0+d'_1\eps$ with $D'_0$ an integer ($D'_0\ge 2$, see appendix A of \refr{Somogyi:2008fc}), $a_1$ and $a_2$ are {\em finite} as $\eps\to 0$, which is necessary for \eqnss{eq:Sr}{eq:CarSr} to make sense as subtraction terms in $d=4$ dimensions.

Finally, we remark that using \eqn{eq:C2} to simplify \eqn{eq:IcCSFF}, we find that the second term of the latter equation vanishes, and thus we have (compare \eqns{eq:IcCSFF}{eq:IcCSIF})
\beq
\IcCS{FF}(\eps) = \IcCS{IF}(\eps)\,.
\eeq

~\\
\noindent{\bf Final results.}
Having determined the $a_n$, we can write the integrated soft-type counterterms explicitly. Here we limit ourselves to recalling the pole parts of the combinations
\beq
\IcS{ik}{FF}(Y;\eps) + \IcCS{FF}(\eps)\,,
\quad
\IcS{ai}{IF}(Y;\eps) 
	+ \frac{1}{2}\left[\IcCS{FF}(\eps)+\IcCS{IF}(\eps)\right]\,,
\quad\mbox{and}\quad
\IcS{ab}{II}(\eps) + \IcCS{IF}(\eps)\,,
\eeq
as it is these combinations which finally enter the insertion operator (see \sect{sec:intA1} below). We find that the pole parts are independent of $y_0$ and $d'_0$ and read
\beq
\IcS{ik}{FF}(Y;\eps) + \IcCS{FF}(\eps) =
\IcS{ai}{IF}(Y;\eps) 
	+ \frac{1}{2}\left[\IcCS{FF}(\eps)+\IcCS{IF}(\eps)\right] =
\frac{1}{\eps}\ln Y + \Oe{0}
\label{eq:TcSpoles}
\eeq
and
\beq
\IcS{ab}{II}(\eps) + \IcCS{IF}(\eps) = 0
\label{eq:TcSIIpoles}
\eeq
exactly. (Notice that $2B(-\eps,1-\eps)=B(-\eps,-\eps)$, so that the right hand sides of \eqns{eq:IcCSIF}{eq:IcSabII} coincide up to the overall sign.) The finite part of \eqn{eq:TcSpoles}, with $d'_0=3-3\eps$, is given in \appx{app:IcSfin}.


\section{Implementation of the subtraction scheme}
\label{sec:subimp}

The unpolarized fully differential real-emission cross section with two incoming partons $a$ and $b$ multiplied by the jet function reads
\beq
\bsp
\dsig{R}_{ab}(p_a,p_b) J_{m+1} &= {\cal N}\sum_{\{m+1\}}
	\frac{1}{\Phi(p_a\cdot p_b)}\PS{m+1}(\mom{};p_a+p_b)
	\frac{1}{\omega(f_a)\omega(f_b)}
\\&\times
	\frac{1}{S_{\{m+1\}}} \SME{m+1,ab}{0}{\momab{}} 
	J_{m+1}(\momab{})\,.
\label{eq:dsigRfull}
\esp
\eeq
Here ${\cal N}$ denotes all QCD-independent factors, the factor $1/(\omega(f_a)\omega(f_b))$ accounts for the average over the polarizations and colours of the initial partons and $\Phi(p_a\cdot p_b)$ is the flux factor, which fulfills the following scaling property
\beq
\Phi(\eta p_a\cdot p_b) = \eta\, \Phi(p_a\cdot p_b)\,.
\eeq
Finally, $\sum_{\{m+1\}}$ denotes a sum over the different subprocesses that contribute and $S_{\{m+1\}}$ is the Bose symmetry factor for identical particles in the final state. 

%
%

\subsection{The approximate cross section}
\label{sec:approx_xsec}

In order to compute the NLO cross section, we write it as in \eqn{eq:subtract}, with the approximate cross section $\dsiga{R}{}_{ab}(p_a,p_b)$ given by
\beq
\bsp
\dsiga{R}{}_{ab}(p_a,p_b) J_m &= {\cal N}\sum_{\{m+1\}}
	\frac{1}{\Phi(p_a\cdot p_b)}\PS{m+1}(\mom{};p_a+p_b)
	\frac{1}{\omega(f_a)\omega(f_b)}
\\&\times
	\frac{1}{S_{\{m+1\}}}\cA_{1}\SME{m+1,ab}{0}{\momab{}} 
	\otimes J_{m}\,,
\label{eq:dsigRAfull}
\esp
\eeq
where we define
\beq
\bsp
&\cA_{1}\SME{m+1,ab}{0}{\momab{}}\otimes J_m = 
	\sum_{r\in F}\Bigg[\sum_{\substack{i\in F \\ i\ne r}}
	\frac{1}{2}\cC{ir}{FF}(\momab{})J_m(\momabt{m}{(ir)})
\\&\qquad +
	\sum_{a\in I}\cC{ar}{IF}(\momab{})J_m(\momabtt{m}{(ar)}{ar}{b})
	+ \Bigg(\cS{r}{}(\momab{})
\\&\qquad
	-\sum_{\substack{i\in F \\ i\ne r}}\cC{ir}{FF}\cS{r}{}(\momab{})
	-\sum_{a\in I}\cC{ar}{IF}\cS{r}{}(\momab{})\Bigg)J_m(\momabt{m}{(r)})\Bigg]\,.
\esp
\label{eq:cA1SME}
\eeq
As previously, $F$ and $I$ denote the set of final- and initial-state partons respectively.

We note that the singly-unresolved approximate cross section defined in \eqn{eq:dsigRAfull} above is fully local, i.e.\ all azimuthal and colour correlations are properly taken into account. This means in particular that we can test numerically the convergence of $\dsiga{R}{}_{ab}J_m$ to the real-emission cross section $\dsig{R}_{ab}J_{m+1}$ in any singly-unresolved limit. As a check of the proposed scheme, we have examined the process $gg\to 4g$, where non-trivial azimuthal and colour correlations are both present. By generating sequences of phase space points tending to a particular limit, we have confirmed numerically that
\beq
R\equiv\frac{\dsiga{R}{}_{gg}J_3}{\dsig{R}_{gg}J_{4}}\to 1
\label{eq:Rdef}
\eeq
in all one-parton unresolved limits. This is illustrated in \fig{fig:limits}, where the scatter plots show 100 sequences of 30 points each, starting from random phase space points and converging to a given limit. One sequence of points is highlighted in each case for transparency.
\FIGURE[t]{
\label{fig:limits}
\includegraphics[scale=0.4]{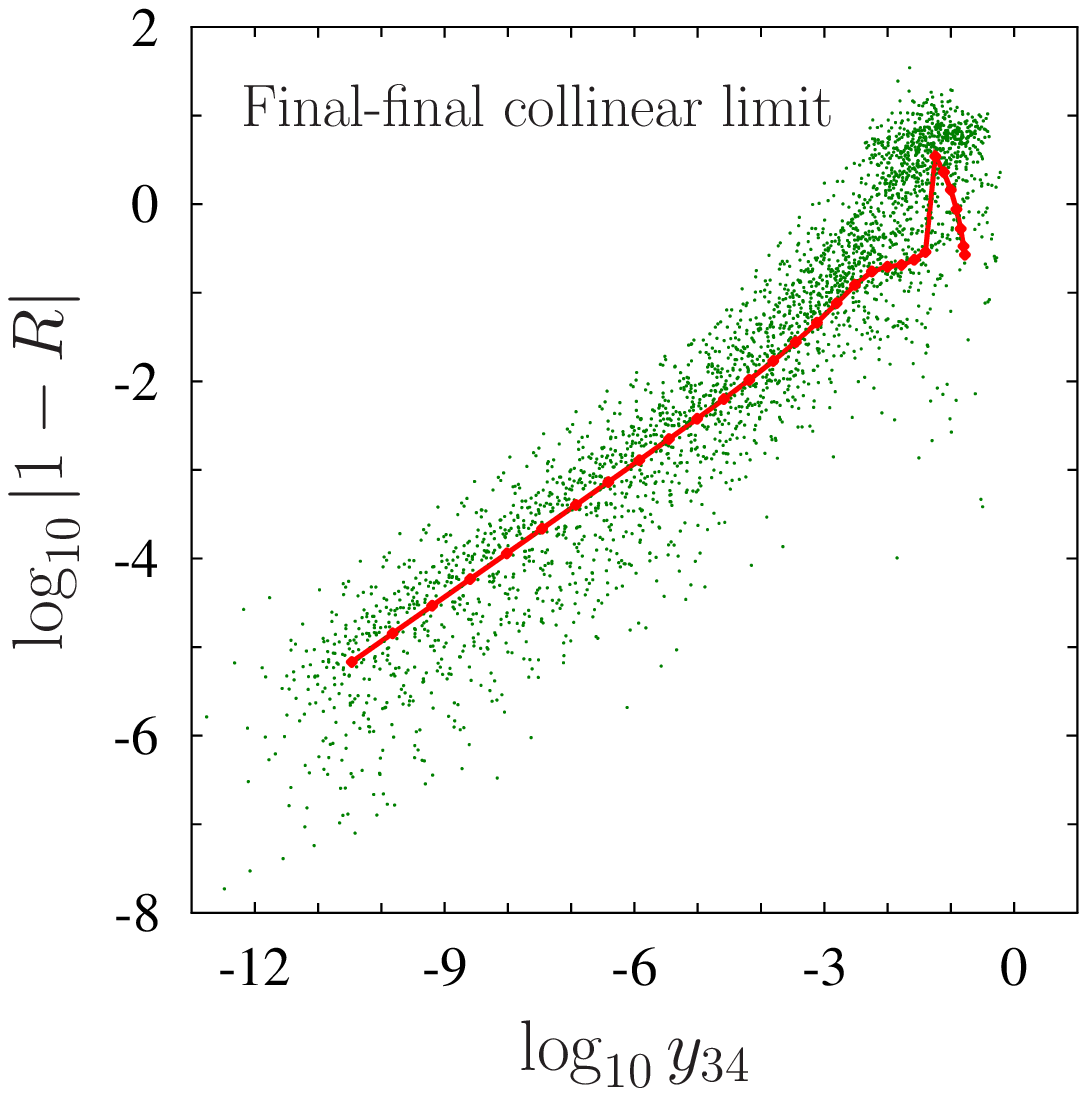}
\includegraphics[scale=0.4]{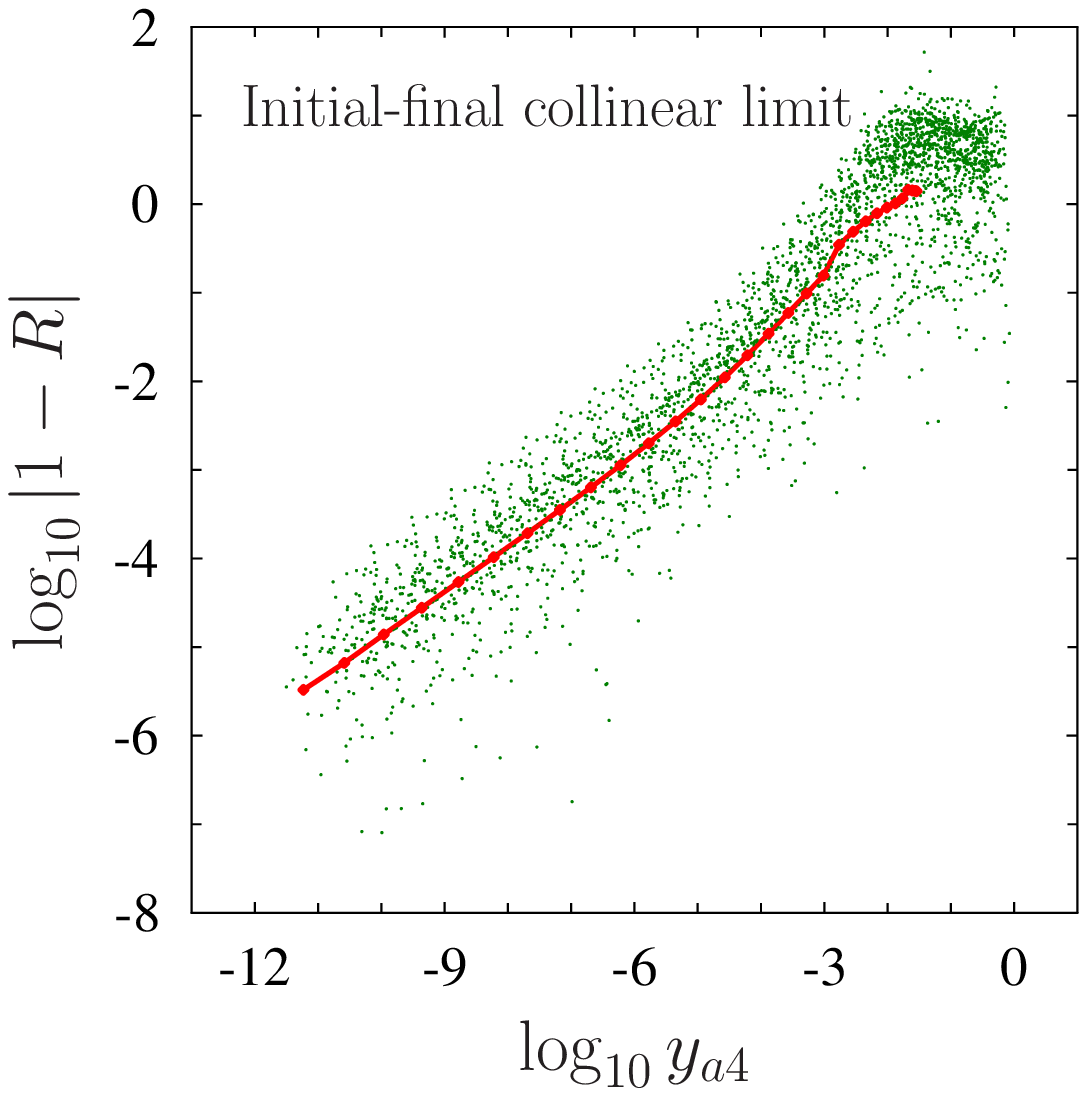}
\includegraphics[scale=0.4]{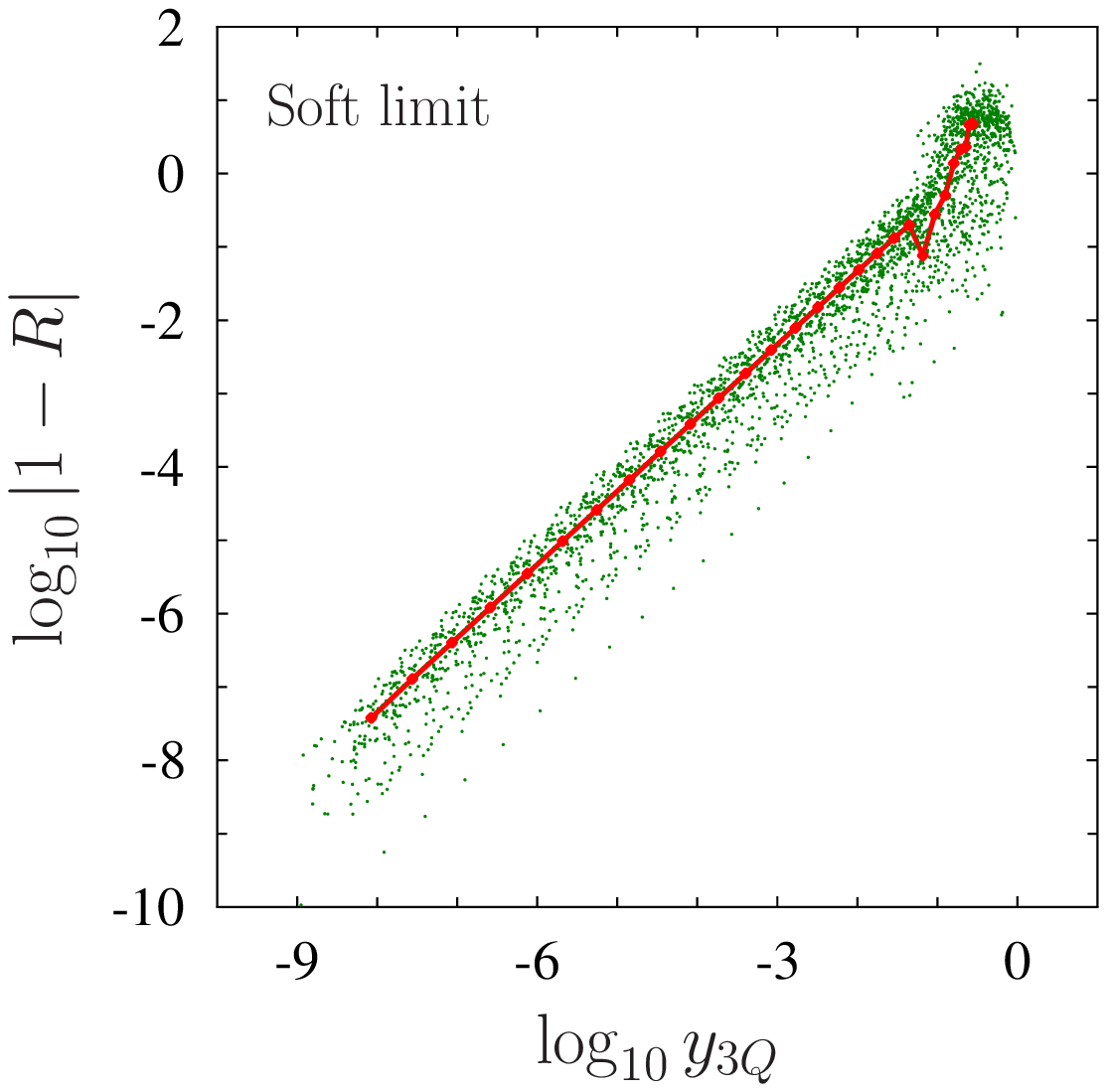}
\caption{
Scatter plots of 100 sequences of 30 points each demonstrating the convergence of the approximate cross section to the real-emission cross section in the process $gg\to 4g$. The ratio $R$ is defined in \eqn{eq:Rdef}. We have set $\alpha_0=y'_0=y_0=1$ and $D_0=D'_0=3$. From left to right the plots show the $\bC{34}$ final-final collinear limit, the $\bC{a4}$ initial-final collinear limit and the $\bS{3}$ soft limit respectively. The labels of the two incoming gluons are $a$ and $b$, the final-state gluons are labeled $1$ through $4$.}
}

%
%

\subsection{The integrated approximate cross section}
\label{sec:intA1}

In order to compute the integral of the approximate cross section, let us write it as follows
\beq
\dsiga{R}{}_{ab}(p_a,p_b) J_m =
	\dsiga{R}{i}_{ab}(p_a,p_b) J_m
	+\dsiga{R}{ii}_{ab}(p_a,p_b) J_m
	+\dsiga{R}{iii}_{ab}(p_a,p_b) J_m\,,
\eeq
where the three terms correspond to the final-final collinear, initial-final collinear and soft-type terms respectively.

To evaluate $\int_{m+1}\dsiga{R}{i}_{ab}(p_a,p_b)J_m$, we use the phase space factorization property of \eqn{eq:PSfactir} and perform the integral to find 
\beq
\bsp
\int_{m+1}\dsiga{R}{i}_{ab}(p_a,p_b)J_m &=
	\int_m {\cal N} \sum_{\{m+1\}}\frac{1}{\Phi(p_a\cdot p_b)}
	\PS{m}(\momt{(ir)}_m;p_a+p_b)\frac{1}{\omega(f_a)\omega(f_b)}
\\&\times
	\frac{1}{S_{\{m+1\}}}\sum_{i\in F}\sum_{\substack{r\in F \\ r\ne i}}
	\frac{\as}{2\pi}S_\eps\left(\frac{\mu^2}{Q^2}\right)^\eps
	\frac{1}{2}\,\IcC{ir}{FF}(y_{\wti{ir}Q};\eps)\,\bT_{ir}^2
\\&\times
	\SME{m,ab}{0}{\momabt{m}{(ir)}}J_m(\momabt{m}{(ir)})\,.
\esp
\label{eq:intAi_1}
\eeq
This expression is not yet in the form of an $m$-parton contribution times a factor. In order to rewrite \eqn{eq:intAi_1} in such a form, we still have to perform the counting of the symmetry factors in going form $m$ to $m+1$ partons. This counting was performed in \refr{Catani:1996vz} (see their eq.\ (7.19)) and here we only recall the final result
\beq
\bsp
\sum_{\{m+1\}}\frac{1}{S_{\{m+1\}}}\sum_{i\in F}\sum_{\substack{r\in F \\ r\ne i}}
	\frac{1}{2}\ldots & =
	\sum_{\{m\}}\!\strut^{(i)}\frac{1}{S_{\{m\}}}
	\Bigg(\sum_{\substack{\wti{ir}\in F \\ f_{\wti{ir}}=q}}\ldots
	+\sum_{\substack{\wti{ir}\in F \\ f_{\wti{ir}}=\qb}}\ldots
	+\frac{1}{2}\sum_{\substack{\wti{ir}\in F \\ f_{\wti{ir}}=g}}\ldots\Bigg)
\\& +
	\sum_{\{m\}}\!\strut^{(ii)}\frac{1}{S_{\{m\}}}\Nf
	\sum_{\substack{\wti{ir}\in F \\ f_{\wti{ir}}=g}}\ldots\,,
\esp
\label{eq:FFcounting}
\eeq
where the upper indices on the sums indicate that we may go from $m$ to $m+1$ parton final states by (i) adding a gluon or (ii) exchanging a gluon for a quark-antiquark pair. Substituting \eqn{eq:FFcounting} into \eqn{eq:intAi_1} we find
\beq
\bsp
\int_{m+1}\dsiga{R}{i}_{ab}(p_a,p_b)J_m &=
	\int_m {\cal N} \sum_{\{m\}}\frac{1}{\Phi(p_a\cdot p_b)}
	\PS{m}(\momt{(ir)}_m;p_a+p_b)\frac{1}{\omega(f_a)\omega(f_b)}
\\&\times
	\frac{1}{S_{\{m\}}}\sum_{\wti{ir}\in F}
	\frac{\as}{2\pi}S_\eps\left(\frac{\mu^2}{Q^2}\right)^\eps
	\IcC{\wti{ir}}{F}(y_{\wti{ir}Q};\eps)\,\bT_{ir}^2
\\&\times
	\SME{m,ab}{0}{\momabt{m}{(ir)}}J_m(\momabt{m}{(ir)})\,,
\esp
\label{eq:intAi_2}
\eeq
where we have defined 
\bal
\IcC{q}{F}(x;\eps) &=\IcC{\qb}{F}(x;\eps)\equiv \IcC{qg}{FF}(x;\eps)\,,
\\[.5em]
\IcC{g}{F}(x;\eps) &\equiv \frac{1}{2}\IcC{gg}{FF}(x;\eps) + \Nf \IcC{q\qb}{FF}(x;\eps)\,.
\eal
For further reference, we note that the pole parts of $\IcC{i}{F}(x;\eps)$ may be written in the unified form
\beq
\IcC{i}{F}(x;\eps) = \frac{1}{\bT_i^2}\left(\frac{\bT_i^2}{\eps^2}
	+\gamma_i \frac{1}{\eps}\right) - \frac{2}{\eps}\ln(x) + \Oe{0}
\eeq
with the usual flavour constants
\beq
\gamma_q = \gamma_\qb = \frac{3}{2}\CF\,,
\quad
\gamma_g = \frac{11}{6}\CA - \frac{2}{3}\TR\Nf\,.
\eeq

Next we consider the integral $\int_{m+1}\dsiga{R}{ii}_{ab}(p_a,p_b)J_m$. Using the phase space convolution property of \eqn{eq:PSconvar} and performing the integral we find
\beq
\bsp
\int_{m+1}\dsiga{R}{ii}_{ab}(p_a,p_b)J_m &=
	\int_m {\cal N} \sum_{\{m+1\}}\int_0^1 \rd x\,\frac{1}{\Phi(xp_a\cdot p_b)}
	\PS{m}(\momt{(ar)}_m;x p_a+p_b)
\\&\times
	\frac{1}{\omega(f_{ar})\omega(f_b)}
	\frac{1}{S_{\{m+1\}}}\sum_{r\in F}
	\frac{\as}{2\pi}S_\eps\left(\frac{\mu^2}{Q^2}\right)^\eps
	\IcC{a(ar)}{I}(x;\eps)\,\bT_{ar}^2
\\&\times
	\SME{m,(ar)b}{0}{\momabtx{m}{(ar)}{x}{}}J_m(\momabtx{m}{(ar)}{x}{})
	+(a \leftrightarrow b)\,.
\esp
\label{eq:intAii_1}
\eeq
To get \eqn{eq:intAii_1} in the form of an $m$-parton contribution times a factor we again have to perform the counting of symmetry factors. This counting was performed in \refr{Catani:1996vz} (see their eq.\ (8.18)) and is anyway trivial in this case,
\beq
\sum_{\{m+1\}}\frac{1}{S_{\{m+1\}}}\sum_{r\in F}\ldots = 
	\sum_{f_r} \sum_{\{m\}}\frac{1}{S_{\{m\}}}\ldots\,.
\label{eq:IFcounting}
\eeq
Using the flavour sum rules, we can rewrite the sum over the flavours of $r$ as a sum over the flavours of the initial-state parton $ar$ ($c=f_{ar})$, and \eqn{eq:intAii_1} becomes
\beq
\bsp
\int_{m+1}\dsiga{R}{ii}_{ab}(p_a,p_b)J_m &=
	\int_m {\cal N} \sum_{\{m\}}\int_0^1 \rd x\,\frac{1}{\Phi(xp_a\cdot p_b)}
	\sum_{c}\PS{m}(\momt{(c)}_m;x p_a+p_b)
\\&\times
	\frac{1}{\omega(f_c)\omega(f_b)}
	\frac{1}{S_{\{m\}}}
	\frac{\as}{2\pi}S_\eps\left(\frac{\mu^2}{Q^2}\right)^\eps
	\IcC{ac}{I}(x;\eps)\,\bT_{c}^2
\\&\times
	\SME{m,cb}{0}{\momabtx{m}{(c)}{x}{}}J_m(\momabtx{m}{(c)}{x}{})
	+(a \leftrightarrow b)\,.
\esp
\label{eq:intAii_2}
\eeq

Finally we turn to the evaluation of $\int_{m+1}\dsiga{R}{iii}_{ab}(p_a,p_b)J_m$. Inserting the phase space factorization of \eqn{eq:PSfactr} and performing the integral we obtain
\beq
\bsp
&\int_{m+1}\dsiga{R}{iii}_{ab}(p_a,p_b)J_m =
\\&\qquad =
	\int_m {\cal N} \sum_{\{m+1\}}\frac{1}{\Phi(p_a\cdot p_b)}
	\PS{m}(\momt{(r)}_m;p_a+p_b)\frac{1}{\omega(f_a)\omega(f_b)}
	\frac{1}{S_{\{m+1\}}}\frac{\as}{2\pi}S_\eps\left(\frac{\mu^2}{Q^2}\right)^\eps
\\&\qquad\times
	\sum_{r\in F}
	\Bigg[\sum_{i\in F}\sum_{\substack{k\in F \\ k\ne i}}
	\IcS{ik}{FF}(\Y{i}{k};\eps)
	\bra{m,ab}{(0)}{(\momabt{m}{(r)})}\bT_i\bT_k\ket{m,ab}{(0)}{(\momabt{m}{(r)})}
\\&\qquad\qquad +
	2\sum_{i\in F}\IcS{ai}{IF}(Y_{a\ti{i},Q};\eps)
	\bra{m,ab}{(0)}{(\momabt{m}{(r)})}\bT_a\bT_i\ket{m,ab}{(0)}{(\momabt{m}{(r)})}
\\&\qquad\qquad +
	2\sum_{i\in F}\IcS{bi}{IF}(Y_{b\ti{i},Q};\eps)
	\bra{m,ab}{(0)}{(\momabt{m}{(r)})}\bT_b\bT_i\ket{m,ab}{(0)}{(\momabt{m}{(r)})}
\\&\qquad\qquad +
	2\,\IcS{ab}{II}(\eps)
	\bra{m,ab}{(0)}{(\momabt{m}{(r)})}\bT_a\bT_b\ket{m,ab}{(0)}{(\momabt{m}{(r)})}
	\rule[-1.1em]{0pt}{2.2em}
\\&\qquad\qquad -
	\sum_{i \in F}\IcCS{FF}(\eps)\,\bT_i^2\SME{m,ab}{0}{\momabt{m}{(r)}}
	-\IcCS{IF}(\eps)\,\bT_a^2\SME{m,ab}{0}{\momabt{m}{(r)}}
\\&\qquad\qquad\qquad -
	\IcCS{IF}(\eps)\,\bT_b^2\SME{m,ab}{0}{\momabt{m}{(r)}}
\Bigg]J_m(\momabt{m}{(r)})\,.
\esp
\label{eq:intAiii_1}
\eeq 
Using colour conservation (\eqn{eq:colourcons}), we may combine the integrated soft and soft-collinear terms and write \eqn{eq:intAiii_1} in the following form
\beq
\bsp
&\int_{m+1}\dsiga{R}{iii}_{ab}(p_a,p_b)J_m =
\\&\qquad =
	\int_m {\cal N} \sum_{\{m+1\}}\frac{1}{\Phi(p_a\cdot p_b)}
	\PS{m}(\momt{(r)}_m;p_a+p_b)\frac{1}{\omega(f_a)\omega(f_b)}
	\frac{1}{S_{\{m+1\}}}\frac{\as}{2\pi}S_\eps\left(\frac{\mu^2}{Q^2}\right)^\eps
\\&\qquad\times
	\sum_{r\in F}
	\Bigg[\sum_{i\in F}\sum_{\substack{k\in F \\ k\ne i}}
	\TcS{ik}{FF}(\Y{i}{k};\eps)
	\bra{m,ab}{(0)}{(\momabt{m}{(r)})}\bT_i\bT_k\ket{m,ab}{(0)}{(\momabt{m}{(r)})}
\\&\qquad\qquad +
	2\sum_{i\in F}\TcS{ai}{IF}(Y_{a\ti{i},Q};\eps)
	\bra{m,ab}{(0)}{(\momabt{m}{(r)})}\bT_a\bT_i\ket{m,ab}{(0)}{(\momabt{m}{(r)})}
\\&\qquad\qquad +
	2\sum_{i\in F}\TcS{bi}{IF}(Y_{b\ti{i},Q};\eps)
	\bra{m,ab}{(0)}{(\momabt{m}{(r)})}\bT_b\bT_i\ket{m,ab}{(0)}{(\momabt{m}{(r)})}
\\&\qquad\qquad +
	2\,\TcS{ab}{II}(\eps)
	\bra{m,ab}{(0)}{(\momabt{m}{(r)})}\bT_a\bT_b\ket{m,ab}{(0)}{(\momabt{m}{(r)})}
\Bigg]J_m(\momabt{m}{(r)})\,,
\esp
\label{eq:intAiii_2}
\eeq 
where as claimed, the integrated soft and soft-collinear counterterms appear only in the following combinations:
\bal
\TcS{ik}{FF}(Y;\eps) &\equiv \IcS{ik}{FF}(Y;\eps) + \IcCS{FF}(\eps)\,,
\\[.5em]
\TcS{ai}{IF}(Y;\eps) &\equiv \IcS{ai}{IF}(Y;\eps) 
	+ \frac{1}{2}\left[\IcCS{FF}(\eps)+\IcCS{IF}(\eps)\right]\,,
\\[.5em]
\TcS{ab}{II}(\eps) &\equiv \IcS{ab}{II}(\eps) + \IcCS{IF}(\eps)\,.
\eal
Recall from \sect{sec:cS-type} that in fact $\TcS{ab}{II}(\eps) = 0$ and $\TcS{ik}{FF}(Y;\eps) = \TcS{ai}{IF}(Y;\eps)$. Furthermore, the pole parts of the latter functions are very simple:
\beq
\TcS{ik}{FF}(Y;\eps) = \TcS{ai}{IF}(Y;\eps) = \frac{1}{\eps}\ln Y + \Oe{0}\,.
\eeq

To write \eqn{eq:intAiii_2} in the form of an $m$-parton contribution times a factor, we use the relations between the $m$- and $(m+1)$-parton symmetry factors, $S_{\{m+1\}} = (m_g + 1) S_{\{m\}}$, and note that for each $r$ of the $m_g+1$ final-state gluons the expression in the square brackets in \eqn{eq:intAiii_2} contributes the same. Therefore,
\beq
\sum_{\{m+1\}}\frac{1}{S_{\{m+1\}}} \sum_{r \in F} \ldots =
\sum_{\{m\}}\frac{1}{S_{\{m\}}} \ldots\,,
\eeq
and we obtain
\beq
\bsp
&\int_{m+1}\dsiga{R}{iii}_{ab}(p_a,p_b)J_m =
\\&\qquad =
	\int_m {\cal N} \sum_{\{m\}}\frac{1}{\Phi(p_a\cdot p_b)}
	\PS{m}(\momt{(r)}_m;p_a+p_b)\frac{1}{\omega(f_a)\omega(f_b)}
	\frac{1}{S_{\{m\}}}\frac{\as}{2\pi}S_\eps\left(\frac{\mu^2}{Q^2}\right)^\eps
\\&\qquad\times
	\Bigg[\sum_{i\in F}\sum_{\substack{k\in F \\ k\ne i}}
	\TcS{ik}{FF}(\Y{i}{k};\eps)
	\bra{m,ab}{(0)}{(\momabt{m}{(r)})}\bT_i\bT_k\ket{m,ab}{(0)}{(\momabt{m}{(r)})}
\\&\qquad\qquad +
	2\sum_{i\in F}\TcS{ai}{IF}(Y_{a\ti{i},Q};\eps)
	\bra{m,ab}{(0)}{(\momabt{m}{(r)})}\bT_a\bT_i\ket{m,ab}{(0)}{(\momabt{m}{(r)})}
\\&\qquad\qquad +
	2\sum_{i\in F}\TcS{bi}{IF}(Y_{b\ti{i},Q};\eps)
	\bra{m,ab}{(0)}{(\momabt{m}{(r)})}\bT_b\bT_i\ket{m,ab}{(0)}{(\momabt{m}{(r)})}
\\&\qquad\qquad +
	2\,\TcS{ab}{II}(\eps)
	\bra{m,ab}{(0)}{(\momabt{m}{(r)})}\bT_a\bT_b\ket{m,ab}{(0)}{(\momabt{m}{(r)})}
\Bigg]J_m(\momabt{m}{(r)})\,.
\esp
\label{eq:intAiii_3}
\eeq 

Collecting eqs.\ (\ref{eq:intAi_2}, \ref{eq:intAii_2}, \ref{eq:intAiii_3}), adding \eqn{eq:dsigC}, and using colour conservation, we find that the sum of the integrated approximate cross section and the collinear counterterm can be written in the form
\beq
\bsp
&\int_{m+1}\dsiga{R}{}_{ab}(p_a,p_b) + \int_m\dsig{C}_{ab}(p_a,p_b;\muF{2}) =
\\&\qquad = 
	\int_m {\cal N}\,
	\sum_{\{m\}}\frac{1}{S_{\{m\}}}
	\int_0^1 \rd x\int_0^1 \rd y\, \frac{1}{\Phi(x\, y\, p_a\cdot p_b)}
	\PS{m}(\mom{}_m;xp_a+yp_b)\frac{1}{\omega(f_c)\omega(f_d)}
\\&\qquad\times
	\sum_{c,d\in I}
	\bra{m,cd}{(0)}{(\mom{}_m,xp_a,yp_b)}
	\bI^{ab,cd}(\mom{}_m;p_a,x;p_b,y;\eps;\muF{2})
	\ket{m,cd}{(0)}{(\mom{}_m,xp_a,yp_b)}\,.
\esp
\eeq
The insertion operator $\bI^{ab,cd}$ depends on the colour charges, momenta and flavours of the QCD partons. Its explicit expression can be written as follows
\beq
\bsp
&\bI^{ab,cd}(\mom{}_m;p_a,x;p_b,y;\eps;\muF{2}) =
	\frac{\as}{2\pi}S_{\eps}\left(\frac{\mu^2}{Q^2}\right)^\eps
\\&\qquad\times
	\Bigg\{\Bigg[\sum_{i\in F}\Bigg[\sum_{\substack{k\in F \\ k\ne i}}
	\Big(-\IcC{i}{F}(y_{iQ};\eps) + \TcS{ik}{FF}(Y_{ik,Q};\eps)\Big)\bT_i\bT_k
\\&\qquad\qquad\qquad+
	\Big(-\IcC{i}{F}(y_{iQ};\eps) + 2\TcS{ai}{IF}(Y_{ai,Q};\eps)\Big)\bT_a\bT_i
\\&\qquad\qquad\qquad+
	\Big(-\IcC{i}{F}(y_{iQ};\eps) + 2\TcS{bi}{IF}(Y_{bi,Q};\eps)\Big)\bT_b\bT_i\Bigg]
\\&\qquad\qquad
	+2\TcS{ab}{II}(\eps)\bT_a\bT_b
	\Bigg]\delta_{ac}\,\delta_{bd}\,\delta(1-x)\delta(1-y)
\\&\qquad\quad +
	\Bigg[\sum_{i\in F}
	\Big[-\IcC{ac}{I}(x;\eps)
	-\frac{1}{\bT_c^2}\Big(\frac{1}{\eps}P^{ac}(x)-K^{ac}_\FS(x)\Big)\Big]\bT_c\bT_i
\\&\qquad\qquad+
	\Big[-\IcC{ac}{I}(x;\eps)
	-\frac{1}{\bT_c^2}\Big(\frac{1}{\eps}P^{ac}(x)-K^{ac}_\FS(x)\Big)\Big]\bT_c\bT_b
	\Bigg]\delta_{bd}\,\delta(1-y)
\\&\qquad\quad +
	\Bigg[\sum_{i\in F}
	\Big[-\IcC{bd}{I}(y;\eps)
	-\frac{1}{\bT_d^2}\Big(\frac{1}{\eps}P^{bd}(y)-K^{bd}_\FS(y)\Big)\Big]\bT_d\bT_i
\\&\qquad\qquad+
	\Big[-\IcC{bd}{I}(y;\eps)
	-\frac{1}{\bT_d^2}\Big(\frac{1}{\eps}P^{bd}(y)-K^{bd}_\FS(y)\Big)\Big]\bT_d\bT_a
	\Bigg]\delta_{ac}\,\delta(1-x)\Bigg\}\,.
\esp
\label{eq:Iabcd}
\eeq
It is not difficult to check that the pole part of $\bI^{ab,cd}$ correctly cancels the pole part of the virtual cross section. Using $\TcS{ab}{II}(\eps) = 0$ and substituting the following identities into \eqn{eq:Iabcd}
\beq
\bsp
\sum_{\substack{i,k\in F \\ i\ne k}}
	\Big(-\IcC{i}{F}(y_{iQ};\eps) + \TcS{ik}{FF}(Y_{ik,Q};\eps)\Big)\bT_i\bT_k &=
	\sum_{\substack{i,k\in F \\ i\ne k}}
	\left[-\frac{1}{\bT_i^2}\left(\frac{\bT_i^2}{\eps^2}+\gamma_i\frac{1}{\eps}\right)
	y_{ik}^{-\eps}\right]\bT_i\bT_k + \Oe{0}\,,
\esp
\eeq
\beq
\Big(-\IcC{i}{F}(y_{iQ};\eps) + 2\TcS{ai}{IF}(Y_{ai,Q};\eps)\Big) =
	-\frac{1}{\bT_i^2}\left(\frac{\bT_i^2}{\eps^2}+\gamma_i\frac{1}{\eps}\right)
	y_{ai}^{-\eps} 
	-\frac{1}{\bT_a^2}\,\frac{\bT_a^2}{\eps^2}\, (y_{ai}^{-\eps}-1) + \Oe{0}\,,
\eeq
\beq
\Big[-\IcC{ac}{I}(x;\eps)
	-\frac{1}{\bT_c^2}\Big(\frac{1}{\eps}P^{ac}(x)-K^{ac}_\FS(x)\Big)\Big] =
	-\frac{1}{\bT_c^2}\left(\frac{\bT_c^2}{\eps^2} + \gamma_c \frac{1}{\eps}\right)
	\delta_{ac}\,\delta(1-x) + \Oe{0}\,,
\eeq
we find that the pole structure of \eqn{eq:Iabcd} is exactly as in \eqns{eq:I1dsigRR}{eq:Ieps}. Therefore the sum $\int_1\dsiga{R}{}_{ab}(p_a,p_b)+\dsig{C}_{ab}(p_a,p_b;\muF{2})$ cancels all the singularities in the virtual contribution $\dsig{V}_{ab}(p_a,p_b)$. The cancellation of poles is a strong check on the correctness of the proposed scheme.


\section{Conclusions}
\label{sec:conclusions}

Extending any NLO subtraction algorithm to NNLO accuracy in a process-independent way is non-trivial because the integrated singly-unresolved approximate cross section may fail to have a universal collinear limit. The subtraction scheme in \refr{Somogyi:2006cz} leads to an approximate cross section that does not suffer from this problem and indeed that scheme can be included as part of an NNLO subtraction algorithm without any changes \cite{Somogyi:2006da,Somogyi:2006db}. However, that scheme is defined only for processes with no coloured particles in the initial state.

In this paper, we have presented the generalization of the subtraction algorithm of \refr{Somogyi:2006cz} to processes with hadronic initial states. By matching the known factorization formulae for the collinear and soft limits of QCD squared matrix elements and by carefully extending the matched expression over the full phase space, we have explicitly defined a singly-unresolved approximate cross section $\dsiga{R}{}$ that is completely general (process- and observable-independent) and fully local (i.e.\ all azimuthal and colour correlations are properly taken into account). Furthermore, the integrated approximate cross section obeys universal factorization properties in the collinear and soft limits. It is then possible to build a singly-unresolved approximation to the integrated approximate cross section $\int_1\dsiga{R}{}$ in a process-independent fashion, which is necessary for regularizing the real-virtual contribution in an NNLO computation.

We emphasize that in order to define an approximate cross section whose integrated form obeys universal factorization in the singly-unresolved IR limits, we had to consider a `non-minimal' extension of the (soft and soft-collinear) limit formulae over the phase space.

We have performed the integration of the approximate cross section and found that its pole structure is exactly the same, but with opposite sign, as that of the one-loop squared matrix element. Thus the integrated approximate cross section correctly cancels the poles of the virtual correction. As a further check, we have examined numerically the local convergence of $\dsiga{R}{}_{ab}J_m$ to the real-radiation cross section, $\dsig{R}_{ab}J_{m+1}$, in the process $gg\to 4g$ and found that their ratio tends to unity in any singly-unresolved limit.

All analytic formulae, relevant for constructing a numerical program to compute cross sections at NLO accuracy using the new subtraction algorithm have been presented, and we anticipate that any such implementation will be usable as part of an NNLO calculation without any modifications, once the scheme is defined at NNLO accuracy. Setting up an extension of the present algorithm at NNLO seems straightforward conceptually, but it nevertheless poses a major technical challenge and is therefore left for later work.


\section*{Acknowledgments}

I thank Z.~Tr\'ocs\'anyi for useful discussions and comments on the manuscript. I am also grateful to V.~Del Duca and T.~Gehrmann for their comments on the manuscript. This research was supported in part by the Swiss National Science Foundation (SNF) under contract 200020-117602 and by the Hungarian Scientific Research Fund grant OTKA K-60432.


\appendix


\section{The integrated subtraction terms}
\label{app:intsubterms}

In this appendix we give the finite parts of the various integrated subtraction terms.  We set the exponents $d_0$ and $d'_0$ to $d_0=d'_0=3-3\eps$. We choose to consider this particular value for the reasons explained in app.\ A of \refr{Somogyi:2008fc}.

%
%

\subsection{The integrated collinear subtraction terms}
\label{app:IcCfin}


\subsubsection{Final-final collinear}
\label{app:IcCFFfin}

We denote the finite, $\Oe{0}$ part of the integrated final-final collinear subtraction terms as $\finite\IcC{ir}{FF}(x;\eps)$. For $d_0=3-3\eps$ we find
\beq
\bsp
&\qquad
\ts{\finite\IcC{qg}{FF}(x;\eps) =
-\frac{3 \alpha_0^4}{8 (x-1)}-\frac{3 \alpha_0^4}{8}+\frac{3 \alpha_0^3}{2
   (x-1)}-\frac{\alpha_0^3}{3 (x-1)^2}+\frac{11 \alpha_0^3}{6}-\frac{13
   \alpha_0^2}{6 (x-1)}+\frac{13 \alpha_0^2}{12
   (x-1)^2}-}
\\ &
\ts{\frac{\alpha_0^2}{6 (x-1)^3}-\frac{41 \alpha_0^2}{12}+\frac{5
   \alpha_0}{6 (x-1)}-\frac{7 \alpha_0}{6 (x-1)^2}+\frac{5 \alpha_0}{6
   (x-1)^3}+\frac{2 \alpha_0}{3 (x-1)^4}+\frac{13
   \alpha_0}{6}+\Big(\frac{8}{3}+\frac{3}{2 (x-1)}-\frac{1}{3
   (x-1)^2}-}
\\ &
\ts{\frac{1}{3 (x-1)^3}+\frac{3}{2 (x-1)^4}+\frac{8}{3 (x-1)^5}\Big)
   H(0;\alpha_0)+\Big(-\frac{17}{3}-\frac{3}{2 (x-1)}+\frac{1}{3
   (x-1)^2}+\frac{1}{3 (x-1)^3}-\frac{3}{2 (x-1)^4}-}
\\ &
\ts{\frac{8}{3 (x-1)^5}\Big)
   H(0;x)+\Big(\frac{2}{(x-1)^5}-2\Big) H(0;\alpha_0)
   H(1;x)+\Big(-\frac{\alpha_0^4}{2 (x-1)}+\frac{\alpha_0^4}{2}+\frac{2
   \alpha_0^3}{x-1}-\frac{2 \alpha_0^3}{3 (x-1)^2}-}
\\ &
\ts{\frac{8
   \alpha_0^3}{3}-\frac{3 \alpha_0^2}{x-1}+\frac{2
   \alpha_0^2}{(x-1)^2}-\frac{\alpha_0^2}{(x-1)^3}+6 \alpha_0^2+\frac{2
   \alpha_0}{x-1}-\frac{2 \alpha_0}{(x-1)^2}+\frac{2
   \alpha_0}{(x-1)^3}-\frac{2 \alpha_0}{(x-1)^4}-8 \alpha_0+\frac{3}{2
   (x-1)}-}
\\ &
\ts{\frac{1}{3 (x-1)^2}-\frac{1}{3 (x-1)^3}+\frac{3}{2
   (x-1)^4}+\frac{8}{3 (x-1)^5}+\frac{25}{6}\Big)
   H(c_1(\alpha_0);x)+4
   H(0,0;x)+\Big(\frac{2}{(x-1)^5}-}
\\ &
\ts{2\Big)
   H(0,c_1(\alpha_0);x)+\Big(2-\frac{2}{(x-1)^5}\Big)
   H(1,0;x)+\Big(\frac{2}{(x-1)^5}-2\Big)
   H(1,c_1(\alpha_0);x)-}
\\ &
\ts{\frac{2
   H(c_1(\alpha_0),c_1(\alpha_0);x)}{(x-1)^5}-\frac{\pi
   ^2}{3 (x-1)^5}-\frac{\pi ^2}{6}+\frac{7}{2}}\,,
\esp
\label{eq:FinCFFqg}
\eeq
and
\beq
\bsp
&\qquad
\ts{\finite\IcC{q\qb}{FF}(x;\eps) =
\frac{\TR}{\CA}\Big[
-\frac{2 \alpha_0^6}{3 (x \alpha_0-2 \alpha_0-x)}+\frac{2 \alpha_0^5}{3
   (x-2)}+\frac{10 \alpha_0^5}{3 (x \alpha_0-2 \alpha_0-x)}-\frac{5
   \alpha_0^4}{2 (x-2)}+\frac{\alpha_0^4}{6 (x-1)}-}
\\ &
\ts{\frac{20 \alpha_0^4}{3
   (x \alpha_0-2 \alpha_0-x)}+\frac{5 \alpha_0^4}{3
   (x-2)^2}+\frac{\alpha_0^4}{6}+\frac{10 \alpha_0^3}{3 (x-2)}-\frac{2
   \alpha_0^3}{3 (x-1)}+\frac{20 \alpha_0^3}{3 (x \alpha_0-2
   \alpha_0-x)}-\frac{40 \alpha_0^3}{9 (x-2)^2}+\frac{2 \alpha_0^3}{9
   (x-1)^2}+}
\\ &
\ts{\frac{40 \alpha_0^3}{9 (x-2)^3}-\frac{8 \alpha_0^3}{9}-\frac{5
   \alpha_0^2}{3 (x-2)}+\frac{\alpha_0^2}{x-1}-\frac{10 \alpha_0^2}{3 (x
   \alpha_0-2 \alpha_0-x)}+\frac{10 \alpha_0^2}{3 (x-2)^2}-\frac{2
   \alpha_0^2}{3 (x-1)^2}-\frac{20 \alpha_0^2}{3
   (x-2)^3}+\frac{\alpha_0^2}{3 (x-1)^3}+}
\\ &
\ts{\frac{40 \alpha_0^2}{3 (x-2)^4}+2
   \alpha_0^2-\frac{2 \alpha_0}{3 (x-1)}+\frac{2 \alpha_0}{3 (x \alpha_0-2
   \alpha_0-x)}+\frac{2 \alpha_0}{3 (x-1)^2}-\frac{2 \alpha_0}{3
   (x-1)^3}+\frac{2 \alpha_0}{3 (x-1)^4}+\frac{160 \alpha_0}{3
   (x-2)^5}-\frac{8 \alpha_0}{3}+}
\\ &
\ts{\Big(\frac{2}{3
   (x-1)^5}+\frac{2}{3}+\frac{160}{3 (x-2)^5}+\frac{320}{3 (x-2)^6}\Big)
   H(0;\alpha_0)+\Big(-\frac{2}{3 (x-1)^5}+\frac{2}{3}-\frac{160}{3
   (x-2)^5}-}
\\ &
\ts{\frac{320}{3 (x-2)^6}\Big) H(0;x)+\frac{2
   H(c_1(\alpha_0);x)}{3 (x-1)^5}+\Big(\frac{160}{3
   (x-2)^5}+\frac{320}{3 (x-2)^6}\Big)
   H(c_2(\alpha_0);x)+\Big(\frac{160}{3(x-2)^5}+}
\\ &
\ts{\frac{320}{3(x-2)^6}\Big)\ln (2)-\frac{10}{9}\Big]}\,.
\esp
\label{eq:FinCFFqq}
\eeq
The other two integrated collinear functions are not independent of \eqns{eq:FinCFFqg}{eq:FinCFFqq}. Firstly, we have the trivial relationship that $\IcC{qg}{FF}(x;\eps)$ and $\IcC{gq}{FF}(x;\eps)$ are equal. Secondly, the integrated gluon-gluon splitting function satisfies
\beq
\IcC{gg}{FF}(x;\eps) = 2\IcC{qg}{FF}(x;\eps) 
	- (1-\eps)\frac{\CA}{\TR}\IcC{q\qb}{FF}(x;\eps)\,,
\eeq
and so for the finite part we have
\beq
\finite\IcC{gg}{FF}(x;\eps) = 2\finite\IcC{qg}{FF}(x;\eps) 
	-\frac{\CA}{\TR}\finite\IcC{q\qb}{FF}(x;\eps)-\frac{2}{3}\,.
\eeq
The $H$ functions appearing in \eqns{eq:FinCFFqg}{eq:FinCFFqq} denote one- and two-dimensional harmonic polylogarithms \cite{Remiddi:1999ew,Gehrmann:2000zt}. Those involving $c_1(\alpha_0)$ or $c_2(\alpha_0)$ were defined in \refr{Aglietti:2008fe} by an extension of the standard basis for {\em 2dHPL's} (see \appx{app:HPLS} below). However, all {\em HPL's} and {\em 2dHPL's} that appear above can be written in terms of logarithms and dilogarithms. We present the explicit expressions in \appx{app:HPLS}.

For $\alpha_0=1$, the expressions above simplify quite significantly, and we find
\beq
\bsp
&\qquad
\ts{\finite\IcC{qg}{FF}(x;\eps;\alpha_0=1) =
-\frac{5}{24
   (x-1)}-\frac{5}{12 (x-1)^2}+\frac{2}{3 (x-1)^3}+\frac{2}{3
   (x-1)^4}+\Big(-\frac{17}{3}-\frac{3}{2 (x-1)}+}
\\ &
\ts{\frac{1}{3(x-1)^2}+\frac{1}{3 (x-1)^3}-\frac{3}{2 (x-1)^4}-\frac{8}{3 (x-1)^5}\Big)
   \ln (x)+2 \ln ^2(x)+\Big(2-\frac{2}{(x-1)^5}\Big) \text{Li}_2(1-x)-\frac{\pi ^2}{2}+\frac{89}{24}}\,,
\esp
\label{eq:FinCFFqga01}
\eeq
and
\beq
\bsp
&\qquad
\ts{\finite\IcC{q\qb}{FF}(x;\eps;\alpha_0=1) =
\frac{\TR}{\CA}\Big[
-\frac{1}{6 (x-2)}-\frac{1}{6
   (x-1)}+\frac{5}{9 (x-2)^2}+\frac{2}{9 (x-1)^2}-\frac{20}{9
   (x-2)^3}-}
\\ &
\ts{\frac{1}{3 (x-1)^3}+\frac{40}{3 (x-2)^4}+\frac{2}{3
   (x-1)^4}+\frac{160}{3 (x-2)^5}+\Big(\frac{160}{3 (x-2)^5}+\frac{320}{3
   (x-2)^6}\Big)\ln (2)+\Big(-\frac{2}{3 (x-1)^5}+\frac{2}{3}-}
\\ &
\ts{\frac{160}{3
   (x-2)^5}-\frac{320}{3 (x-2)^6}\Big) \ln (x)-\frac{5}{2}\Big]}\,.
\esp
\label{eq:FinCFFqqa01}
\eeq
Finally we point out that despite the appearance of factors of $1/(1-x)$ in eqs.\ (\ref{eq:FinCFFqg}, \ref{eq:FinCFFqq}, \ref{eq:FinCFFqga01}) and (\ref{eq:FinCFFqqa01}) above, the integrated collinear functions are finite at $x=1$ as expected, and we have explicitly
\beq
\bsp
\ts{\finite\IcC{qg}{FF}(x=1;\eps)} &=
\ts{\frac{19 \alpha_0^5}{50}-\frac{9 \alpha_0^4}{4}+\frac{49
   \alpha_0^3}{9}-\frac{13 \alpha_0^2}{2}+\frac{3 \alpha_0}{2}
	+\Big(-\frac{2 \alpha_0^5}{5}+\frac{3
   \alpha_0^4}{2}-\frac{4 \alpha_0^3}{3}-2 \alpha_0^2+6\alpha_0}
\\ &-
\ts{\frac{3}{2}\Big) \ln (\alpha_0)-\ln^2(\alpha_0)
	-\frac{\pi ^2}{2}+\frac{7}{2}}\,,
\esp
\label{eq:FinCFFqgx1}
\eeq
and
\beq
\bsp
\ts{\finite\IcC{q\qb}{FF}(x=1;\eps)} &=
\ts{\frac{\TR}{\CA}\Big[
-\frac{2 \alpha_0^5}{15}+\alpha_0^4-\frac{34 \alpha_0^3}{9}+11
   \alpha_0^2-36 \alpha_0+\frac{2 \ln (\alpha_0)}{3}+\frac{160}{3} \ln
   (\alpha_0+1)}
\\ &+
\ts{\frac{64}{3 (\alpha_0+1)}-\frac{202}{9}\Big]}\,.
\esp
\label{eq:FinCFFqqx1}
\eeq
For both $\alpha_0=1$ and $x=1$, we have simply
\bal
\ts{\finite\IcC{qg}{FF}(x=1;\eps;\alpha_0=1)} &= 
	\ts{\frac{1867}{900}-\frac{\pi ^2}{2}}\,,
\label{eq:FinCFFqga01x1}
\\
\ts{\finite\IcC{q\qb}{FF}(x=1;\eps;\alpha_0=1)} &=
	\ts{\frac{\TR}{\CA}\Big[
-\frac{1786}{45}+\frac{160 \ln (2)}{3}\Big]}\,.
\label{eq:FinCFFqqa01x1}
\eal
In passing we note that the forms of these functions as given in \eqns{eq:FinCFFqg}{eq:FinCFFqq} (or in \eqns{eq:FinCFFqga01}{eq:FinCFFqqa01} for $\alpha_0=1$) are not particularly well suited for direct numerical evaluation very close to $x=1$ because of issues of numerical stability. However, since the functions are smooth at $x=1$, it is straightforward to develop simple approximations to any desired accuracy around this one point, e.g.\ most simply by Taylor-expanding around $x=1$, with the leading terms given in \eqns{eq:FinCFFqgx1}{eq:FinCFFqqx1} (respectively in \eqns{eq:FinCFFqga01x1}{eq:FinCFFqqa01x1} for $\alpha_0=1$).


\subsubsection{Initial-final collinear}
\label{app:IcCIFfin}

The finite part of the integrated initial-final collinear subtraction terms is denoted by $\finite\IcC{ab}{I}(x;\eps)$. For generic $y'_0$, we find
\bal
\finite\IcC{qg}{I}(x;\eps) =&
	\frac{\CF}{\CA}\bigg\{
	\frac{1+(1-x)^2}{x}
	\Big[\ln(1-x) + \ln(1-x)\Theta(x-(1-y'_0)) 
\nt\\ &\qquad
	+ \ln y'_0\Theta((1-y'_0)-x)\Big] + x\bigg\}\,,
\\[.5em]
\finite\IcC{gq}{I}(x;\eps) =&
	\frac{\TR}{\CF}\bigg\{\left[x^2+(1-x)^2\right]
	\Big[\ln(1-x) + \ln(1-x)\Theta(x-(1-y'_0)) 
\nt\\ &\qquad
	+ \ln y'_0\Theta((1-y'_0)-x)\Big] + 2x(1-x)\bigg\}\,,
\\[.5em]
\finite\IcC{qq}{I}(x;\eps) =&
	2\left(\frac{\ln(1-x)}{1-x}\right)_+ + 2\left(\frac{\ln(1-x)}{1-x}\right)_{1-y'_0}
	+\frac{2\ln y'_0}{1-x}\Theta((1-y'_0)-x)
\nt\\ &
	-\Big\{(1+x)\Big[\ln(1-x) + \ln(1-x)\Theta(x-(1-y'_0)) 
	+ \ln y'_0\Theta((1-y'_0)-x)\Big] 
\nt\\ &\qquad
	-1+x\Big\} + \delta(1-x)\left(-\frac{\pi^2}{6}+\ln^2 y'_0\right)\,,
\\[.5em]
\finite\IcC{gg}{I}(x;\eps) =&
	2\left(\frac{\ln(1-x)}{1-x}\right)_+ + 2\left(\frac{\ln(1-x)}{1-x}\right)_{1-y'_0}
	+\frac{2\ln y'_0}{1-x}\Theta((1-y'_0)-x)
\nt\\ &
	+2\left(\frac{1-x}{x}-1+x(1-x)\right)\Big[\ln(1-x) + \ln(1-x)\Theta(x-(1-y'_0)) 
\nt\\ &\qquad
	+ \ln y'_0\Theta((1-y'_0)-x)\Big] 
	+ \delta(1-x)\left(-\frac{\pi^2}{6}+\ln^2 y'_0\right)\,.
\eal
Above the `$1-y'_0$' prescription is defined by its action on a generic test function $g(x)$ as follows
\beq
\int \rd x\,g(x)[f(x)]_{1-y'_0} \equiv \int_{1-y'_0}^1 \rd x\,[g(x)-g(1)]f(x)\,.
\eeq

These integrals were first computed in \refr{Nagy:2003tz}. However, our notation is sufficiently different form the one employed in \cite{Nagy:2003tz} that a direct comparison is not entirely straightforward. Therefore, we give the expression for $\IcC{ab}{I}(x;\eps)$ in terms of the various functions introduced in \refr{Nagy:2003tz}. We find
\beq
\IcC{ab}{I}(x;\eps) = -\frac{1}{\eps}P^{ab}(x)+\widetilde{K}^{ab}(x,y'_0)
	+\overline{K}^{ab}(x,y'_0) + P^{ab}(x)\ln x 
	+ \delta^{ab}\delta(1-x){\cal V}_a(y'_0,\eps) + \Oe{1}\,,
\eeq 
where $P^{ab}(x)$ are the four-dimensional Altarelli--Parisi probabilities given in \eqnss{eq:Psplitqg}{eq:Psplitgg}, while $\widetilde{K}^{ab}(x,y'_0)$, $\overline{K}^{ab}(x,y'_0)$ and ${\cal V}_a(y'_0,\eps)$ are all defined in \cite{Nagy:2003tz} (note also that our $y'_0$ is the $\alpha$ of \refr{Nagy:2003tz}).

Finally, we remind the reader that for $y'_0=1$, the $\IcC{ab}{I}(x;\eps)$ functions  become identical (up to a colour factor) to the $\widetilde{{\cal V}}^{a,b}(x;\eps)$ functions of \refr{Catani:1996vz}. The precise correspondence is given by
\beq
\IcC{ab}{I}(x;\eps;y'_0=1)\bT_{b}^2 = \widetilde{{\cal V}}^{a,b}(x;\eps)\,.
\eeq

%
%

\subsection{The integrated soft-type subtraction terms}
\label{app:IcSfin}

For completeness, we give the explicit solution of \eqns{eq:C1}{eq:C2} determining $a_1$ and $a_2$ for general $y_0$ and $d'_0$:
\bal
\nt
a_1 &=
	\Big\{B_{y_0}(1-2\eps,d'_0)\Big[B_{y_0}(2-2\eps,d'_0)-B_{y_0}(2-2\eps,d'_0+\shalf)\Big]
\\&\qquad
	-B_{y_0}(3-2\eps,d'_0)\Big[B_{y_0}(-2\eps,d'_0)-B_{y_0}(-2\eps,d'_0+\shalf)\Big]\Big\}
\nt\\&\times
	\Big\{B_{y_0}(2-2\eps,d'_0)\Big[B_{y_0}(2-2\eps,d'_0)-B_{y_0}(2-2\eps,d'_0+\shalf)\Big]
\nt\\&\qquad
	-B_{y_0}(3-2\eps,d'_0)\Big[B_{y_0}(1-2\eps,d'_0)-B_{y_0}(1-2\eps,d'_0+\shalf)\Big]\Big\}^{-1}\,,
\label{eq:a1sol}
\\[0.5em]
\nt
a_2 &=
	\Big\{B_{y_0}(2-2\eps,d'_0)\Big[B_{y_0}(-2\eps,d'_0)-B_{y_0}(-2\eps,d'_0+\shalf)\Big]
\\&\qquad
	-B_{y_0}(1-2\eps,d'_0)\Big[B_{y_0}(1-2\eps,d'_0)-B_{y_0}(1-2\eps,d'_0+\shalf)\Big]\Big\}
\nt\\&\times
	\Big\{B_{y_0}(2-2\eps,d'_0)\Big[B_{y_0}(2-2\eps,d'_0)-B_{y_0}(2-2\eps,d'_0+\shalf)\Big]
\nt\\&\qquad
	-B_{y_0}(3-2\eps,d'_0)\Big[B_{y_0}(1-2\eps,d'_0)-B_{y_0}(1-2\eps,d'_0+\shalf)\Big]\Big\}^{-1}\,.
\label{eq:a2sol}
\eal
Note that the difference $B_{y_0}(-2\eps,d'_0)-B_{y_0}(-2\eps,d'_0+\shalf)$ is finite in $\eps$. For $d'_0=3-3\eps$, the solution reads
\beq
\bsp
&\qquad
\ts{a_1 =
\Big[-\frac{y_0^7}{60}+\frac{8 y_0^6}{45}-\frac{209
   y_0^5}{300}+\frac{31 y_0^4}{30}-\frac{1027
   y_0^3}{1890}+\frac{4 y_0^2}{63}-\frac{4
   y_0}{63}+\sqrt{1-y_0} \Big(\frac{4 y_0^7}{675}-\frac{331
   y_0^6}{4725}+\frac{781 y_0^5}{1575}-}
\\ &
\ts{\frac{881
   y_0^4}{945}+\frac{506 y_0^3}{945}-\frac{2
   y_0^2}{63}+\frac{4 y_0}{63}\Big)+\Big(-\frac{2
   y_0^5}{5}+y_0^4-\frac{2 y_0^3}{3}\Big) \ln
   \Big(\sqrt{1-y_0}+1\Big)+\Big(\frac{2
   y_0^5}{5}-y_0^4+}
\\ &
\ts{\frac{2 y_0^3}{3}\Big) \ln
   (2)\Big]\Big/\Big[-\frac{y_0^8}{240}+\frac{y_0^7}{30}-\frac{7
   y_0^6}{60}+\frac{47 y_0^5}{210}-\frac{61
   y_0^4}{252}+\frac{26 y_0^3}{189}-\frac{2
   y_0^2}{63}+\sqrt{1-y_0} \Big(\frac{y_0^8}{630}-\frac{29
   y_0^7}{1890}+\frac{17 y_0^6}{270}-}
\\ &
\ts{\frac{13
   y_0^5}{90}+\frac{5 y_0^4}{27}-\frac{23
   y_0^3}{189}+\frac{2 y_0^2}{63}\Big)\Big] + \Oe{1}}\,,
\esp
\eeq
and
\beq
\bsp
&\qquad
\ts{a_2 =
\Big[\frac{y_0^6}{72}-\frac{y_0^5}{6}+\frac{41
   y_0^4}{60}-\frac{299 y_0^3}{315}+\frac{26
   y_0^2}{105}+\frac{2 y_0}{7}+\sqrt{1-y_0}
   \Big(-\frac{y_0^6}{210}+\frac{13 y_0^5}{210}-\frac{65
   y_0^4}{126}+\frac{61 y_0^3}{63}-}
\\ &
\ts{\frac{41
   y_0^2}{105}-\frac{2
   y_0}{7}\Big)+\Big(\frac{y_0^4}{2}-\frac{4
   y_0^3}{3}+y_0^2\Big) \ln
   \Big(\sqrt{1-y_0}+1\Big)+\Big(-\frac{y_0^4}{2}+\frac{4
   y_0^3}{3}-y_0^2\Big) \ln
   (2)\Big]\Big/\Big[-\frac{y_0^8}{240}+}
\\ &
\ts{\frac{y_0^7}{30}-\frac{7
   y_0^6}{60}+\frac{47 y_0^5}{210}-\frac{61
   y_0^4}{252}+\frac{26 y_0^3}{189}-\frac{2
   y_0^2}{63}+\sqrt{1-y_0} \Big(\frac{y_0^8}{630}-\frac{29
   y_0^7}{1890}+\frac{17 y_0^6}{270}-\frac{13
   y_0^5}{90}+\frac{5 y_0^4}{27}-\frac{23
   y_0^3}{189}+}
\\ &
\ts{\frac{2 y_0^2}{63}\Big)\Big]+\Oe{1}}\,.
\esp
\eeq
As advertised, the solution is finite in $\eps$, and in the expansions above we have kept only the $\Oe{0}$ terms, since they are enough to compute the finite part in the expansion of $\TcS{ik}{FF}(Y;\eps)$ and $\TcS{ai}{IF}(Y;\eps)$.

We note that for $y_0=1$, the above expressions for $a_1$ and $a_2$ simplify to
\bal
\nt
\ts{a_1(y_0=1)} &= \ts{-\frac{3448}{5}+1008 \ln 2 + \Oe{1}
\simeq
	9.09236 + \Oe{1}}\,,
\\[0.5em]
\ts{a_2(y_0=1)} &= \ts{1734-2520 \ln 2 + \Oe{1}
\simeq
	-12.7309 + \Oe{1}}\,.
\eal

Next, we present the finite parts of the integrated soft terms $\TcS{ik}{FF}(Y;\eps)=\TcS{ai}{IF}(Y;\eps)$, denoted by $\finite\TcS{ik}{FF}(Y;\eps)$ and $\finite\TcS{ai}{IF}(Y;\eps)$ respectively. For the specific value of $d'_0=3-3\eps$, we find
\beq
\bsp
&\qquad
\ts{\finite\TcS{ik}{FF}(Y;\eps) = \finite\TcS{ai}{IF}(Y;\eps) = 
-\Big\{2
   \Big[\frac{y_0^{10}}{480}-\frac{y_0^9}{40}+\frac{31
   y_0^8}{225}-\frac{2789 y_0^7}{6300}+\frac{14171
   y_0^6}{16200}-\frac{139 y_0^5}{135}+}
\\ &
\ts{\frac{1313
   y_0^4}{1890}-\frac{17 y_0^3}{63}+\frac{4
   y_0^2}{63}+\sqrt{1-y_0} \Big(-\frac{89
   y_0^{10}}{113400}+\frac{241 y_0^9}{22680}-\frac{173
   y_0^8}{2520}+\frac{526 y_0^7}{2025}-\frac{8584
   y_0^6}{14175}+\frac{761 y_0^5}{945}-\frac{1103
   y_0^4}{1890}+}
\\ &
\ts{\frac{5 y_0^3}{21}-\frac{4
   y_0^2}{63}\Big)+\Big(\frac{y_0^8}{120}-\frac{y_0^7}{15
   }+\frac{7
   y_0^6}{30}-\frac{y_0^5}{3}+\frac{y_0^4}{6}\Big) \ln
   \Big(\sqrt{1-y_0}+1\Big)+\Big(-\frac{y_0^8}{120}+\frac{y_0^7}{15}-\frac{7
   y_0^6}{30}+\frac{y_0^5}{3}-}
\\ &
\ts{\frac{y_0^4}{6}\Big) \ln
   (2)\Big] \ln
   (Y)\Big\}\Big/\Big\{-\frac{y_0^8}{240}+\frac{y_0^7}{30}-\frac{7
   y_0^6}{60}+\frac{47 y_0^5}{210}-\frac{61
   y_0^4}{252}+\frac{26 y_0^3}{189}-\frac{2
   y_0^2}{63}+\sqrt{1-y_0} \Big(\frac{y_0^8}{630}-\frac{29
   y_0^7}{1890}+}
\\ &
\ts{\frac{17 y_0^6}{270}-\frac{13
   y_0^5}{90}+\frac{5 y_0^4}{27}-\frac{23
   y_0^3}{189}+\frac{2 y_0^2}{63}\Big)\Big\}
+\Big(4 y_0-y_0^2\Big) \ln Y-\frac{1}{2} \ln^2 Y-2 \ln (y_0) \ln Y-\Li_2(1-Y)}\,.
\esp
\label{eq:StildeF}
\eeq
Setting $y_0=1$ leads to substantial simplification:
\beq
\bsp
&
\ts{\finite\TcS{ik}{FF}(Y;\eps;y_0=1) = \finite\TcS{ai}{IF}(Y;\eps;y_0=1) =}
\\&\qquad =
\ts{	- \Big[\Big(\frac{2516}{15}- 252 \ln 2\Big) \ln Y
	+\frac{1}{2} \ln^2 Y 
	+\Li_2(1-Y)\Big]}\,.
\esp
\eeq

Finally, we remind the reader that $\TcS{ab}{II}(\eps) = 0$, and this is true to all orders in $\eps$, for any $y_0$ and $d'_0$. Therefore we have $\TcS{ik}{FF}(Y=1;\eps)=\TcS{ai}{IF}(Y=1;\eps)=0$ as well, to all orders in $\eps$. Clearly, $\finite\TcS{ik}{FF}(Y;\eps)$ and $\finite\TcS{ai}{IF}(Y;\eps)$ as given in \eqn{eq:StildeF} are indeed zero at $Y=1$.


\section{Explicit expressions for some harmonic polylogarithms}
\label{app:HPLS}

In this appendix we collect all one- and two-dimensional harmonic polylogarithms that appear in the integrated final-final collinear subtraction terms (see \eqns{eq:FinCFFqg}{eq:FinCFFqq}), expressed in terms of logarithms and dilogarithms. The weight one {\em HPL's} are
\bal
H(0;\alpha_0) =& \ln \alpha_0\,,
\\
H(0;x) =& \ln x\,,
\\
H(1;x) =& -\ln(1-x)\,,
\label{eq:HPLw1}
\eal
while the weight two {\em HPL's} read
\bal
H(0,0;x) =& \frac{1}{2}\ln^2 x\,,
\\
H(1,0;x) =& \Li_2(1-x)-\frac{\pi^2}{6}\,.
\eal
The functions involving $c_1(\alpha_0)$ or $c_2(\alpha_0)$ were defined by an extension of the standard basis for {\em 2dHPL's} in \refr{Aglietti:2008fe}. The new basis functions read
\beq
f(c_1(\alpha_0);x) = \frac{1}{x-c_1(\alpha_0)}\,,
\qquad
f(c_2(\alpha_0);x) = \frac{1}{x-c_2(\alpha_0)}\,,
\eeq
with
\beq
c_1(\alpha_0) = \frac{\alpha_0}{\alpha_0-1}\,,
\qquad
c_2(\alpha_0) = \frac{2\alpha_0}{\alpha_0-1}\,.
\eeq
The {\em 2dHPL's} involving $c_1(\alpha_0)$ are
\bal
H(c_1(\alpha_0);x) =& \ln\left(1+\frac{1-\alpha_0}{\alpha_0}x\right)\,,
\\
H(0,c_1(\alpha_0);x) =& -\Li_2\left(-\frac{1-\alpha_0}{\alpha_0}x\right)\,,
\\
H(1,c_1(\alpha_0);x) =&
	-\ln[(1-\alpha_0)(1-x)]\ln\left(1+\frac{1-\alpha_0}{\alpha_0}x\right)
\\&
	+\Li_2(\alpha_0)-\Li_2[1+(1-\alpha_0)x]\,,
\\
H(c_1(\alpha_0),c_1(\alpha_0);x) =& 
	\frac{1}{2}\ln^2\left(1+\frac{1-\alpha_0}{\alpha_0}x\right)\,.
\eal
Finally, the single function involving $c_2(\alpha_0)$ evaluates to
\beq
H(c_2(\alpha_0);x) = \ln\left(1+\frac{1-\alpha_0}{2\alpha_0}x\right)\,.
\eeq

\end{document}